\RequirePackage{fix-cm}

\makeatletter
\def\@cons#1#2{\begingroup\let\@elt\relax\xdef#1{\ifx#1\relax\else#1\fi\@elt #2}\endgroup}
\makeatother

\documentclass[epj]{svjour}

\usepackage[utf8]{inputenc}
\usepackage[USenglish]{babel}
\usepackage{epsfig}
\usepackage{csquotes}
\usepackage{graphicx}
\usepackage{hyperref}
\usepackage{amsmath}
\smartqed   \usepackage[capitalise,noabbrev]{cleveref}
\usepackage{subcaption}
\usepackage{float}
\usepackage{color}
\usepackage{xspace}
\newcommand{\mt}[1]{\textnormal{#1}}
\newcommand{\mtb}[1]{\textnormal{\textbf{#1}}}
\newcommand{\pandabf}{$\overline{\mtb{P}}$ANDA\xspace}
\newcommand{\panda}{$\overline{\mt{P}}$ANDA\xspace}
\newcommand{\unit}[1]{\, \textnormal{#1}\xspace}
\newcommand{\massunit}{$\,\textnormal{GeV}/c^2$\xspace}
\newcommand{\momentumunit}{$\,\textnormal{GeV}/c$\xspace}
\newcommand{\cascade}{$\Xi^{-}$\xspace}
\newcommand{\anticascade}{$\overline{\Xi}^{+}$\xspace}
\newcommand{\cascasbar}{$\overline{\Xi}\Xi$\xspace}
\newcommand{\cascasbarpinull}{$\overline{\Xi}^+\Xi^-\pi^0$\xspace}
\newcommand{\excitedcascade}{$\Xi^{*-}$\xspace}
\newcommand{\excitedanticascade}{$\overline{\Xi}^{*+}$\xspace}
\newcommand{\excitedcascadefifteen}{$\Xi\left(1530\right)^{-}$\xspace}
\newcommand{\excitedcascadefifteenmin}{$\Xi\left(1530\right)^{-}$\xspace}
\newcommand{\excitedanticascadefifteen}{$\overline{\Xi}\left(1530\right)^{+}$\xspace}

\newcommand{\excitedcascadetwenty}{$\Xi\left(1820\right)^{-}$\xspace}
\newcommand{\excitedanticascadetwenty}{$\overline{\Xi}\left(1820\right)^{+}$\xspace}
\newcommand{\excitedcascadesixteen}{$\Xi\left(1690\right)^{-}$\xspace}
\newcommand{\excitedanticascadesixteen}{$\overline{\Xi}\left(1690\right)^{+}$\xspace}
\newcommand{\pbarp}{$\bar{\textnormal{{p}}}$p\xspace}
\newcommand{\cc}{charge conjugate\xspace}
\newcommand{\lam}{$\Lambda$\xspace}
\newcommand{\alam}{$\bar{\Lambda}$\xspace}
\newcommand{\piminus}{$\pi^{-}$\xspace}
\newcommand{\piplus}{$\pi^{+}$\xspace}
\newcommand{\pinull}{$\pi^0$\xspace}
\newcommand{\kminus}{$\textnormal{K}^{-}$\xspace}
\newcommand{\kplus}{$\textnormal{K}^{+}$\xspace}
\newcommand{\aprot}{$\bar{\textnormal{p}}$\xspace}
\newcommand{\mychannel}{\pbarp $\rightarrow$ \anticascade \excitedcascade}
\newcommand{\mychannelcc}{\pbarp $\rightarrow$ \excitedanticascade \cascade}
\newcommand{\mychannelfscc}{\pbarp $\rightarrow\Xi^-\bar{\Lambda}\mt{K}^+$\xspace}
\newcommand{\channelalbrecht}{\pbarp $\rightarrow$ \cascasbarpinull}

\newcommand{\fs}{$\overline{\Xi}^+\Lambda\mt{K}^-$\xspace}
\newcommand{\fscc}{$\Xi^-\bar{\Lambda}\mt{K}^+$\xspace}
\newcommand{\fsalbrecht}{$\bar{\mt{p}}\mt{p}\pi^+\pi^+\pi^-\pi^-$}

\newcommand{\mychannelfs}{\pbarp $\rightarrow$ \fs}
\newcommand{\decay}[3]{{#1} $\rightarrow$ {#2} + {#3}}
\newcommand{\ctau}{c\tau}
\newcommand{\chisq}{$\chi^2$\xspace}
\newcommand{\DTF}{DecayTreeFitter\xspace}
\begin{document}\sloppy
	\raggedbottom
	\hugehead

\title{\textbf{Study of Excited $\mathbf{\Xi}$ Baryons with the \pandabf Detector}}
%
%
\author{
	G.~Barucca\inst{1} \and 
	F.~Davì\inst{1} \and 
	G.~Lancioni\inst{1} \and 
	P.~Mengucci\inst{1} \and 
	L.~Montalto\inst{1} \and 
	P. P.~Natali\inst{1} \and 
	N.~Paone\inst{1} \and 
	D.~Rinaldi\inst{1} \and 
	L.~Scalise\inst{1} \and 
	B.~Krusche\inst{2}
% ...
}
\institute{
	Università Politecnica delle Marche-Ancona,{ \bf Ancona}, Italy \and 
	Universität Basel,{ \bf Basel}, Switzerland
% ...
}
\date{December 2020}
\abstract{
		The study of baryon excitation spectra provides insight into the inner structure of baryons. So far, 
		most of the world-wide efforts have been directed towards $N^*$ and $\Delta$ spectroscopy.
		Nevertheless, the study of the double and triple strange baryon spectrum provides independent information to the $N^*$ and $\Delta$ spectra.\\
		The future antiproton experiment \panda will provide direct access to final states containing a \cascasbar pair, for which production cross sections up to $\mu\mt{b}$ are expected in \pbarp reactions.
		With a luminosity of $L=10^{31}\unit{cm}^{-2}\mt{s}^{-1}$ in the first phase of the experiment, the expected cross sections correspond to a production rate of $\sim 10^6\, \mt{events}/\mt{day}$.
		With a nearly $4\pi$ detector acceptance, \panda will thus be a hyperon factory.\\
		In this study, reactions of the type \mychannel as well as \mychannelcc with various decay modes are investigated.
		For the exclusive reconstruction of the signal events a full decay tree fit is used, resulting in reconstruction efficiencies between $3\,\%$ and $5\,\%$. This allows high statistics data to be collected within a few weeks of data taking.		
}
\maketitle
	
	\section{Introduction}
	\label{sec:introduction}
	The strong coupling constant $\alpha_s$ increases with decreasing momentum transfer, until at a scale of the proton radius the value of $\alpha_s$ is so large that perturbative methods no longer are applicable.
	Theoretical models used to quantitatively predict hadronic processes in this kinematic regime need to be constrained by experimental data.
	Two classes of approaches are well established.
	One of them is Lattice Quantum Chromodynamics (LQCD) \cite{Wilson1974} which solves the non-perturbative QCD by using numerical simulations. LQCD has given impressive results for hadron spectroscopy \cite{Lin2008,Horsley2011,Padmanath2018} and low-energy physics \cite{aoki2014,della2020,illa2020,aoki2020} during the last decades.
	The other class of approaches  are effective theories that exploit
	the chiral symmetry of the QCD Lagrangian \cite{gasser1985,Weinberg1990,epelbaum2009,petschauer2020,hyodo2020}.
	At low energy, the exchange of hadrons appears to describe the appropriate degrees of freedom for the excitation spectrum and the scattering cross section of baryonic resonances.
	For a deeper insight into the mechanism of non-perturbative QCD the understanding of the excitation pattern of baryons is essential.
	Hadrons are composite particles which have internal degrees of freedom and thus an excitation spectrum.
	This leads to two possibilities to study hadrons in experiments. 
	One possibility is to study reaction dynamics, i.e. the investigation of hadron-hadron interactions and hadron production, while the other is hadron spectroscopy, where the structure of hadrons is investigated.
	There is a long history of calculations of the baryon spectrum within the Constituent Quark Model (CQM) \cite{capstick1986,glozman1998,lring2001}.
	In the CQM the baryon is described as system three quarks or antiquarks which are bound by some confining interaction . 
	Most systematic experimental studies so far have focused on the nucleon excitation spectrum.
	Recently, studies of the $\Delta$ and $N^*$ excited states with the hypercentral Constituent Quark Model (hCQM) \cite{giannini2003,giannini2015,Shah2019,Menapara2020} have been performed .
	In contrast, the knowledge is poor for excited double or triple strange baryon states, also called hyperons.
	Based on the SU(3) flavor symmetry, the $\Xi$ spectrum should contain as many states as the $N^*$ and $\Delta$ spectrum together.\\
	Hyperons are unstable particles and thus unveil more information on their characteristics than nucleons.
	Hence, hyperons, especially their decay, are a powerful tool to address physics problems like the internal structure and fundamental symmetries.\\
	For most hyperons the excitation spectra as well as the ground state properties are still not well understood.	
	Antiproton-proton (\pbarp) induced reactions resulting in a baryon-antibaryon pair provide a good opportunity to access these properties and spectra, since a high fraction of the inelastic \pbarp cross section is associated to final states with a baryon-antibaryon pair together with additional mesons.
	In the \pbarp entrance channel, the production of extra strange mesons is not needed to balance the strangeness in the production of strange or multi-strange baryons.
	In addition, it is possible to directly populate intermediate states, where one hyperon or both hyperons are in an excited state.
	The excited states will predominantly give rise to final states consisting of a baryon-antibaryon pair and one or more mesons, where the produced particles may further decay weakly or electromagnetically.
	If the resonant states in the (anti-)baryon-meson combined system are sufficiently narrow, it will be possible to measure their mass and width directly.
	A partial wave analysis will then give the opportunity to access those observables, e.g. spin and parity quantum numbers, which are otherwise difficult to determine directly.\\
	Comprehensive measurements require next generation experiments.
	For instance, Jefferson Lab recently approved the KLF proposal to construct a $K_L$ beam \cite{Amaryan2020}.
	This facility will be able to produce e.g. an estimated $5.3\cdot 10^6$ \excitedcascadetwenty events within the approved 100 days of beam on target. 	
	Furthermore, the future Antiproton Annihilation in Darmstadt (\panda) experiment located at the FAIR facility will be such an experiment \cite{Erni2009}. 
	It will be a multi-purpose detector to study antiproton-proton induced reaction at beam energies between $1.5\momentumunit$ and $15\momentumunit$.
	Therefore, \panda is well-suited for a comprehensive baryon spectroscopy program in the multi-strange and charm sector.
	The expected cross section for final states containing a \anticascade\cascade pair is on the order of $\mu\mt{b}$ \cite{Musgrave1965}, thus giving the possibility to produce $10^6$ (excited) \cascade events per day, which compares favorably to the $5.3\cdot 10^4$ produced events expected per day at KLF.
	The cross section of the reaction \pbarp$\rightarrow \Omega^-\bar{\Omega}^+$ has never been measured, but is predicted to be $\sigma\simeq2\unit{nb}$ at $p_{\bar{p}}=7\momentumunit$ \cite{Kaidalov1994}.\\
	This work presents a feasibility study for the reconstruction of the reaction \mychannel and its \cc channel with the \panda detector, where $\Xi^*$ denotes the following intermediate resonances:
	\excitedcascadefifteen,
	\excitedcascadesixteen and \excitedcascadetwenty.
	Various decay modes of the resonance states are investigated to study the reconstruction into neutral and charged final state particles, for which the detector might have significantly different performance.
	%
	\section{\pandabf}		
\begin{figure*}[htb]
		\centering
		\includegraphics[width=0.8\textwidth]{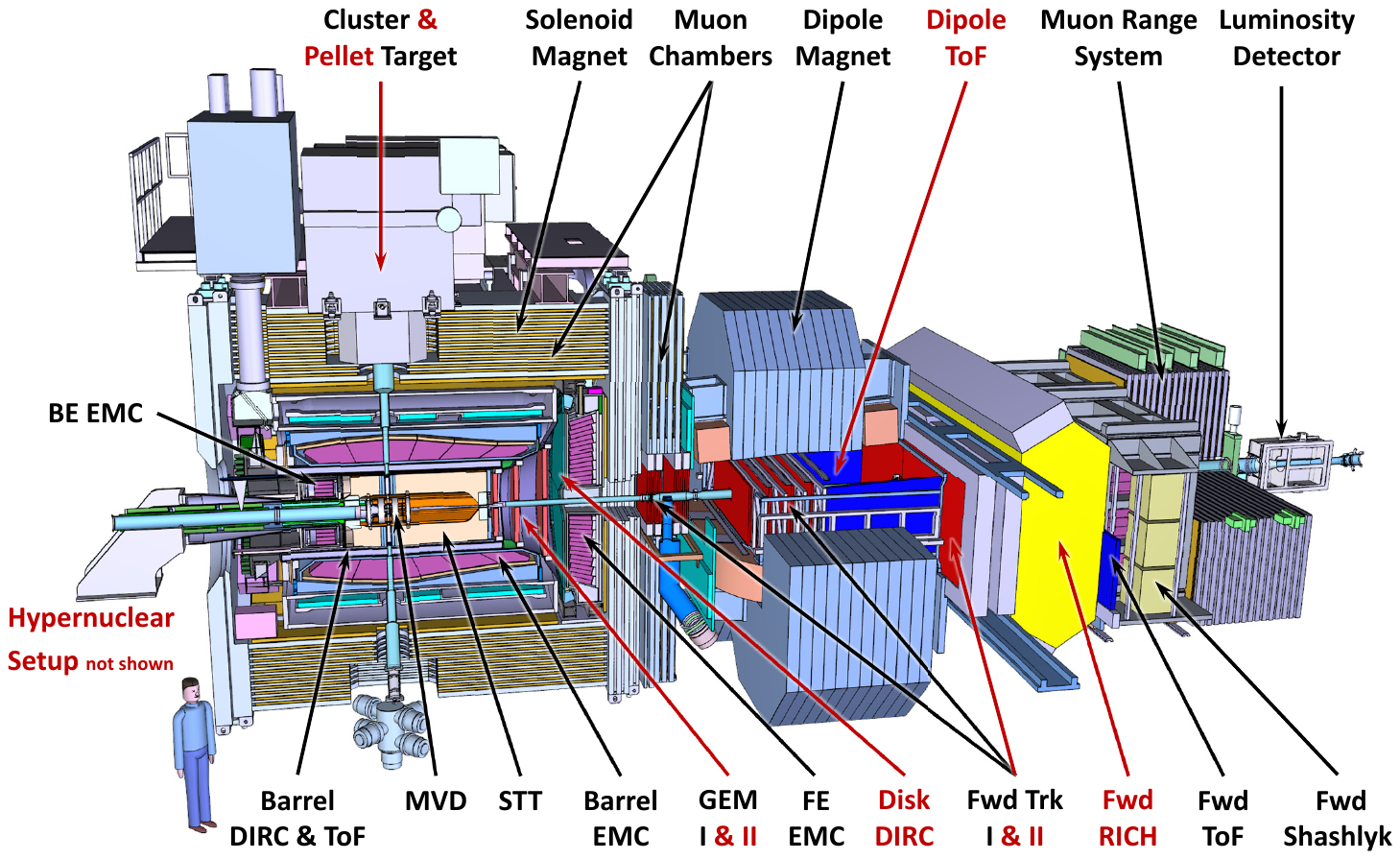}
		\caption{Schematic overview of the \panda detector setup. The components with black labels will be available for the initial configuration of \panda and the components with red labels will be added later. Figure taken from \cite{PANDADetector}.}
		\label{fig:PANDADetector}
	\end{figure*}

	The \panda experiment \cite{Erni2009} will be part of the Facility for Antiproton and Ion Research (FAIR) \cite{FAIR2019}. 
	FAIR is an international accelerator facility for the research with antiprotons and ions, which is currently under construction in Darmstadt, Germany.
	The facility will consist of a system of storage rings.
	One of these storage rings is the High Energy Storage Ring (HESR) which is optimized for high energy antiprotons and will provide a luminosity of about $10^{31}\unit{cm}^{-2}\unit{s}^{-1}$ in the first phase of operation \cite{Schuett2016}.
	HESR can accelerate or decelerate the antiprotons to produce a phase-space cooled beam momentum between $1.5\momentumunit$ and $15\momentumunit$.
	In a later stage a peak luminosity of $2\cdot10^{32}\unit{cm}^{-2}\mt{s}^{-1}$ will be reached \cite{Lehrach2006}.\\
	\newline
	The proposed \panda detector, shown in \cref{fig:PANDADetector}, is a multi-purpose detector and it will be an internal experiment at the HESR.
	It will be composed of two parts, the \textit{Target Spectrometer} (TS) surrounding the interaction point (IP) and the \textit{Forward Spectrometer} (FS).
	This modular design of \panda will lead to almost $4\pi$ geometrical acceptance.\\
	\panda will investigate interactions between the antiproton beam and fixed target protons and/or nuclei.
	Reactions of the antiproton beam on fixed target protons will have a center-of-mass (c.m.) energy between $2.25\unit{GeV}$ and $5.47\unit{GeV}$.
	The target protons will be provided either by a cluster-jet or frozen hydrogen pellets \cite{PandaTDRTarget}.
	In addition, targets of other elements can also be provided for $\bar{\mt{p}}A$ studies.\\
	\panda provides a nearly complete angular coverage, high resolutions for charged and neutral particles as well as a good particle identification.
	The Micro Vertex Detector (MVD) is the innermost part of the tracking system inside the \textit{Target Spectrometer} and uses two different detector technologies: hybrid pixel detectors and double-sided micro-strip detectors \cite{PandaTDRMVD}.
	The main task is to tag events with open charm and strangeness.
	Therefore, the MVD will provide a maximum spatial resolution of $\mu\mt{m}$ perpendicular to and better than $100\,\mu\mt{m}$ along the beam axis.\\
	The main tracking detector for charged particles in the TS is the Straw Tube Tracker (STT), which consists of 4224 single straw tubes arranged in a cylindrical volume around the IP and encloses the MVD \cite{PandaTDRSTT}.
	Together with the MVD and the Gaseous Electron Multiplier (GEM) planes, which are downstream of the STT. 
	The STT is embedded inside the magnetic field of a $2\,\mt{T}$ solenoid \cite{PandaTDRMagnets} giving the possibility to measure the momentum of charged particles. 
	A momentum resolution for charged particles of $\sigma_p/p \sim 1 - 2\,\%$ will be provided by the tracking system of the target spectrometer.\\
	The main charged particle tracking system in the FS is called the Forward Tracker (FTrk) and will consists of three pairs of tracking planes equipped with straw tubes \cite{PandaTDRFTS}.
	The planes will be placed before, inside and behind a $2\unit{T}\cdot\mt{m}$ dipole magnet.
	One of the main tasks is the measurement of particles with low transverse momentum.\\
	A good particle identification (PID) is important for the event reconstruction.
	Therefore, the design of the \panda detector includes PID sub-detectors, i.e. Cherenkov detectors, in particular the Detection of Internal Cherenkov Light (DIRC) \cite{PandaTDRDirc} and the Ring Imaging Cherenkov (RICH) detector, the Barrel Time of Flight (BarrelToF) \cite{PandaTDRBarrelToF} and the Forward Time of Flight (FToF) detector \cite{PandaTDRFToF}, and the Muon Detector System (MDS) \cite{PandaTDRMDS}.\\
	Many channels that will be studied within the physics program of \panda contain photons or electron-positron pairs in the final state.
	The Electromagnetic Calorimeter (EMC) will provide an efficient reconstruction of positron, electron and photons while the background will be suppressed efficiently.
	In the TS the EMC, consisting of the Backward-Endcap EMC (BE EMC), the Barrel EMC and the Forward-Endcap EMC (FE EMC), will be equipped with more than 15,000 PbW$\mt{O}_4$ crystals \cite{PandaTDREMC}.
	In the FS, a shashlyk-type calorimeter is foreseen \cite{PandaTDRFWEMC}.
	The \textit{Forward Spectrometer} will be completed with a Luminosity Detector (LMD) to enable cross section normalization by measuring forward elastically scattered antiprotons in the Coulomb-Nuclear interference region \cite{PandaTDRLumi}.
	\subsubsection*{\textbf{Software Framework}}
	\label{ssec:SoftwareFramework}
	The software framework used to analyze the data is called PandaRoot and is based on ROOT \cite{Brun1997} together with the Virtual Monte Carlo (VMC) package \cite{Hrivnacova2003}.
	The simulation and reconstruction code is implemented within the FairRoot software framework \cite{Al-Turany2012} developed as a common computing structure for all future FAIR experiments \cite{Spataro2011}.
	The detector simulation is handled by VMC and allows the usage of Geant3 \cite{brun1993geant} and Geant4 \cite{Agostinelli2003}.
	Several event generators, i.e. EvtGen \cite{Lange2001EvtGen}, DPM \cite{Capella1994}, UrQMD \cite{bass1998microscopic}, Pythia \cite{sjostrand1997computer} and Fluka \cite{ferrari2005fluka} can be used for the production of signal and background events.
	Subsequently, VMC sends these events to the transport model.
	The detector response after the simulation and propagation of the events is simulated by digitizers.\\
	Charged particle tracks are formed by combining the hits from the tracking detectors.
	For the TS tracking system, the tracking algorithms assume a constant magnetic field and thus helix trajectories for charged particles.
	The Kalman Filter GENFIT \cite{Rauch2015} and the track follower GEANE \cite{Fontana2008} are used to take magnetic field inhomogeneities, energy loss, small angle scattering and the error calculation for the different detector parts into account.
	Up to now, the tracking algorithms use the IP as the origin of the particle track.
	As a consequence, the tracking algorithm has poorer performance for particles emitted far from the IP and thus the standard tracking algorithms do not perform well for the reconstruction of hyperons, which decay with displaced vertices due to their relative long lifetime.
	For this case, an ideal tracking algorithm is used, which groups the hit points into a track based on the generated particle information.\\
	The information of the PID detectors are correlated to the information coming from the reconstructed particles tracks to form charged particles.
	If the particle tracks are not correlated to clusters inside the EMC, neutral candidates are formed.
	For a fast particle identification, algorithms based on Bayesian approaches are implemented \cite{Spataro2011}.
	%
		\section{Event generation and Track Reconstruction $\mathbf{\&}$ Filtering}
	In this section the event generation as well as the procedure for the single track reconstruction and for track filtering are presented.
	\subsection{Event generation}
	\begin{figure}[htb]
		\centering
		\includegraphics[width=0.46\textwidth]{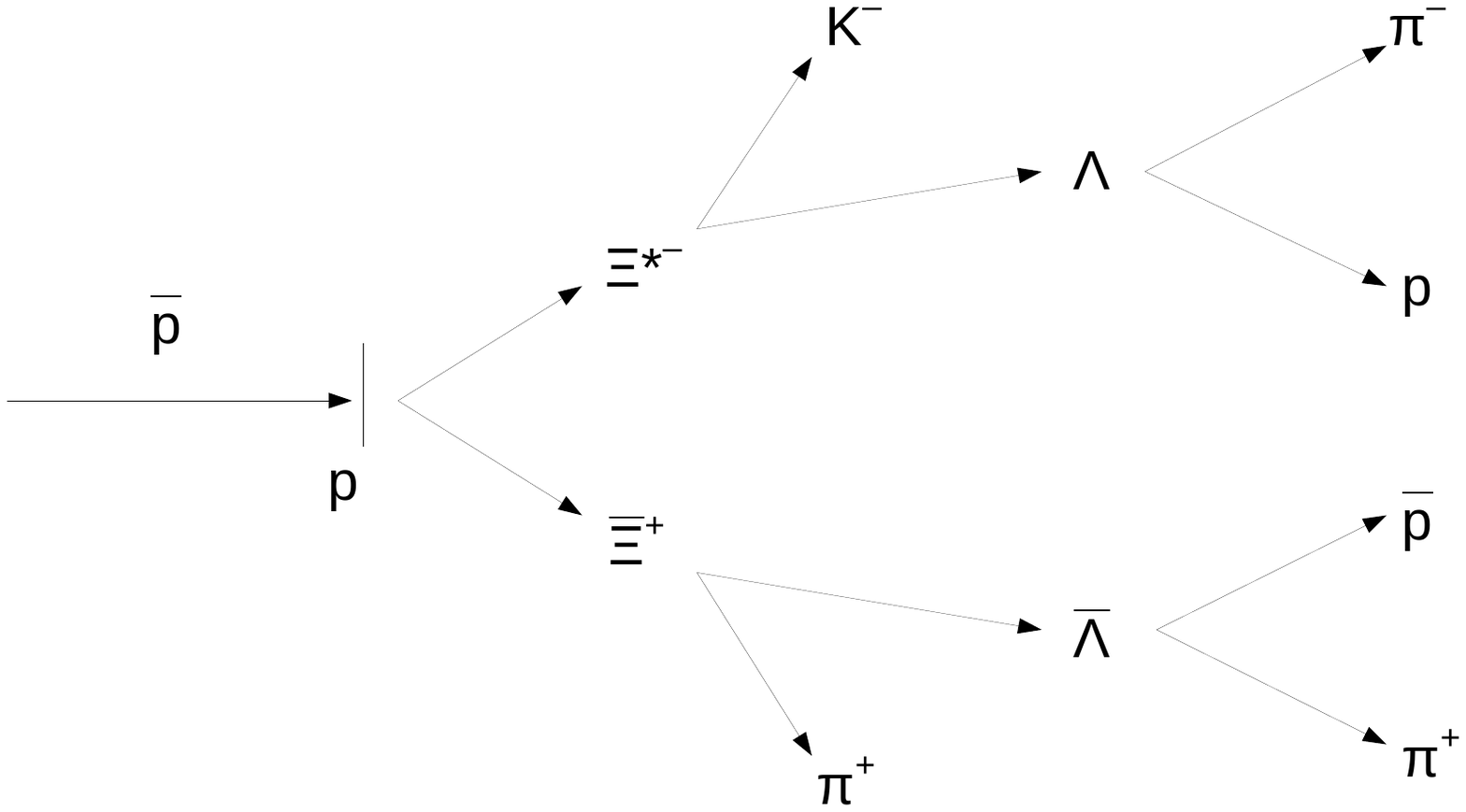}
		\caption{Decay tree for the simulation of \mychannel where $\Xi^*$ decays into \lam\kminus.}
		\label{fig:DecayTree}
	\end{figure}
	In this study, the events to be analyzed, called signal events in the following, were generated with the event generator EvtGen \cite{Lange2001} according to a defined decay chain.
	The decay chain for one of the channels simulated in this work is presented in \cref{fig:DecayTree}.
	The antiproton momentum is chosen to be $p_{\bar{\mt{p}}}=4.6\momentumunit$ corresponding to a c.m. energy of $\sqrt{s}=3.25\unit{GeV}$. 
	The chosen beam momentum allows the population of several resonant states of the $\Xi$ baryon, i.e.
	\excitedcascadefifteen, \excitedcascadesixteen and \excitedcascadetwenty as well as \excitedanticascadefifteen, \excitedanticascadesixteen and \excitedanticascadetwenty.		
	\begin{table}[b]
		\centering
		\caption{Mass and width of the $\Xi$ resonances as implemented for the event generation. The values in parentheses were used for the event generation of the reaction \channelalbrecht.}
		\label{tab:XiResProps}
		\begin{tabular}{lcc}
			\hline
			State & Mass [MeV$/\mt{c}^2$] & $\Gamma$ [MeV$/\mt{c}^2$] \\
			\hline
			\excitedcascadefifteen &  1535 & 9.9 \\
			\excitedcascadesixteen & 1690 & 30 (25) \\
			\excitedcascadetwenty & 1823 & 24 (25) \\
			\hline
		\end{tabular}
	\end{table}
	The properties of the resonant states according to \cite{PDG2018} are summarized in \cref{tab:XiResProps}.
	Different decay channels of the $\Xi$ resonances are investigated:
	\begin{itemize}
		\item \mbox{\excitedcascade$\rightarrow$\lam\kminus},
		\item \excitedcascade$\rightarrow \Xi^- \pi^0$, and 
		\item their \cc channels.
	\end{itemize}  
	The chosen decay channels allow a good test of the reconstruction of far-off vertices (\lam), PID of rare particles (\kplus, \kminus), the reconstruction of composite vertices, \cascade $\rightarrow\pi^-$\lam followed by \lam $\rightarrow\pi^-$p, and also the combination of charged particle information with photon reconstruction (\pinull$\rightarrow \gamma\gamma$).\\
	A non-resonant contribution has been generated in addition to the $\Xi^*$ states mentioned.
	\begin{table*}[t]
		\centering
		\caption{Production and decay branches of the signal events. c.c denotes the \cc.}
		\label{tab:channels}
		\begin{tabular}{lllll}
			\hline
			\pbarp $\rightarrow$ & & & $\rightarrow$ & \fs \\
			\pbarp $\rightarrow$ & \anticascade \excitedcascadesixteen & & $\rightarrow$ & \fs \\
			\pbarp $\rightarrow$ & \anticascade \excitedcascadetwenty& & $\rightarrow$ & \fs\\
			\pbarp $\rightarrow$ & & & $\rightarrow$ & \fscc \\
			\pbarp $\rightarrow$ &\excitedanticascadesixteen \cascade & & $\rightarrow$ & \fscc\\
			\pbarp $\rightarrow$ & \excitedanticascadetwenty \cascade & & $\rightarrow$ & \fscc\\
			\hline
			\pbarp $\rightarrow$ &  & & $\rightarrow$ & \anticascade \cascade \pinull\\
			\pbarp $\rightarrow$ & \anticascade \excitedcascadefifteenmin & (+ c.c.)& $\rightarrow$ & \anticascade \cascade \pinull\\
			\pbarp $\rightarrow$ & \anticascade \excitedcascadesixteen& (+ c.c.)& $\rightarrow$ & \anticascade \cascade \pinull\\
			\pbarp $\rightarrow$ & \anticascade \excitedcascadetwenty& (+ c.c.)& $\rightarrow$ & \anticascade \cascade \pinull\\
			\hline
		\end{tabular}
	\end{table*}
	A full overview of the generated samples is shown in \cref{tab:channels}.
	The ratio between the resonant and non-resonant contribution to the signal events is an assumption based on measured total production cross sections of both excited and ground states of single strange hyperons in \cite{Flaminio1984}.\\
	For each decay mode an isotropic angular distribution is chosen since there are neither experimental data nor theoretical predictions for the reaction \mychannel and its \cc reaction, respectively.
	This simplification ensures that both baryon and anti-baryon are underlying the same detector acceptance.
	In addition, the decay of each resonance is assumed to be isotropic.\\
	Furthermore, the production cross section for \mychannel as well as for \mychannelcc is unknown.
	For the production of \anticascade\cascade in \pbarp collisions at $p=3\momentumunit$ beam momentum a cross section of $\sigma\simeq 2\,\mu\mt{b}$ has been measured \cite{Musgrave1965}.
	In case of single strange hyperons, the comparison of the ground state and the excited state production shows similar cross sections for both species \cite{Flaminio1984}.
	Therefore, the cross section $\sigma$(\mychannel) is assumed to be $1\,\mu\mt{b}$.\\\newline
	Since EvtGen does not take into account the curved trajectory in the magnetic field of the solenoid or the interaction of particles with the detector volume, the propagation of \anticascade and \cascade is passed to Geant4.\\
	The branching ratio for both $\Xi$ baryons to $\Lambda\pi$ is BR($\Xi \rightarrow \Lambda \pi$)$=99.98\,\%$.
	In contrast, \lam as well as \alam have various decay modes with a significant branching ratios.
	Since this study focuses on \decay{\lam}{p}{\piminus} and \decay{\alam}{\aprot}{\piplus} the corresponding branching ratio ($BR=63.4\,\%$) is set to $100\,\%$.
	The final results have been scaled by the correct branching ratios for further calculations.
	\subsection{Track Reconstruction and Filtering}
	\label{ssec:trackfilter}
	A characteristic feature of ground state hyperons is their long decay time, so that they can propagate several centimeters before they decay.
	The lifetimes ($\ctau$) of the \lam and $\Xi$ is are $7.89$ and $4.91\unit{cm}$, respectively \cite{PDG2018}.
	This implies, that their daughter particles are not produced close to the interaction point.
	As mentioned in \cref{ssec:SoftwareFramework}, the tracking algorithms in PandaRoot assume particles to come from the IP meaning that the implemented algorithms are not able to reconstructed the charged final state particles of the reactions to be studied.
	Since no pattern recognition algorithm was available that takes into account particles that decay away from the IP, we used an ideal pattern recognition algorithm instead.
	As a consequence, also particles leaving only one hit in any sub-detector will be reconstructed.
	To simulate a more realistic condition, a track filter is used to reject those tracks with a low hit multiplicity in the tracking detectors.
	In the following, only those charged final state particles are further considered if they leave at least four hits in one of inner tracking detectors (MVD, STT or GEM).
	This selection criterion is motivated by the helix trajectory of a charged particle in a homogeneous magnetic field. 
	Consider f.e. the case the particle is moving along the $z$-axis.
	In that case, the projection of the trajectory onto the $x$-$y$-plane is a circle which can be defined by three hit points inside the detector part. 
	A fourth hit is then a confirmation of the track hypothesis.\\
	The ideal pattern recognition algorithm takes a relative momentum smearing of $5\,\%$ into account.
	Subsequently a Kalman Filter based track fit is applied, which reduces the relative momentum smearing to $\sim1\,\%$.
	%
	\section{The Decay Tree Fit}
	
	\label{sec:DecayTreeFit}
	
	In this section an overview on the method to perform a least-squares fit of a full decay chain is presented.
	For further information the reader can consult \cite{Hulsbergen2005}.\\
	The presented least-squares fit allows a simultaneous extraction of all parameters in a decay chain.
	This method has been developed for the data analysis at the BaBar experiment \cite{Hulsbergen2005}.
	It uses a parameterization in terms of vertex position, momentum and decay time of a particle.\\
	The parameterization of the decay tree is chosen as followed:
	\begin{itemize}
		\item Final state particles are represented by their momentum vector ($p_x, p_y, p_z$), respectively. The mass of the final state particle is assigned by the particle hypothesis set in the decay tree. 
		\item Intermediate state are modeled by a four-momentum vector ($p_x, p_y, p_z, E$) and a decay vertex position($x, y, z$).
		In case the intermediate state is not the initial particle, also the decay time $\theta \equiv l/\left|\vec{p}\right|$, where $l$ is decay length, is used as parameter.
	\end{itemize}
	Furthermore, two types of constraints have to be distinguished: the internal constraints, i.e. vertex constraint and momentum conservation constraint, to remove redundant degrees of freedom, and the external constraint constituted by the reconstructed final state particles.
	The degrees of freedom of the decay tree are formed by the vertex positions and momenta of all involved particles.\\
	The constraints described above are the minimal set of constraints necessary to fit the decay tree starting with the reconstructed final state particles.
	In addition, other constraints, i.e. constraining the mass of composites and the four-momentum of the head, are implemented.
	In principle, missing particles could also be included, if this does not mean that the decay tree is kinematically under-constrained.\\
	The order in which the constraints are applied has an impact on the sum of the \chisq contributions, but with one exception: if all applied constraints are linear, the sum of the \chisq contributions is not affected by the order of the constraints.
	Based on this, the external constraints are applied first, followed by all four-momentum conservation constraints. In the last step geometric constraints as well as mass constraints are applied.\\
	In general, the decay tree fit is repeated until the total \chisq reaches a stable value. In each iteration the parameters are initialized with the results of the previous iteration.
	In contrast, the covariance matrix is reset for each iteration to its original value.
	%
	\section{Event Reconstruction}
	
	\subsection[$\bar{\mt{p}}\mt{p}\rightarrow \bar{\Xi}^+ \Lambda \mt{K}^-$ + c.c] {$\mathbf{\bar{\textbf{p}}\textbf{p}\rightarrow \bar{\Xi}^+ \Lambda \textbf{K}^-}$ + c.c.}
	In this study, in total about 10 million signal events of the reactions \mychannelfs and \mychannelfscc have been analyzed, containing $40\,\%$ \excitedcascadesixteen (\excitedanticascadesixteen), $40\,\%$ \excitedcascadetwenty (\excitedanticascadetwenty), and $20\,\%$ continuum.
	%
	\subsubsection*{\textbf{Final State Particles}}
	\label{sssec:FinalStateSelection}
\begin{figure}[t]
		\begin{subfigure}[b]{0.49\textwidth}
			\centering
			\includegraphics[width=0.8\textwidth]{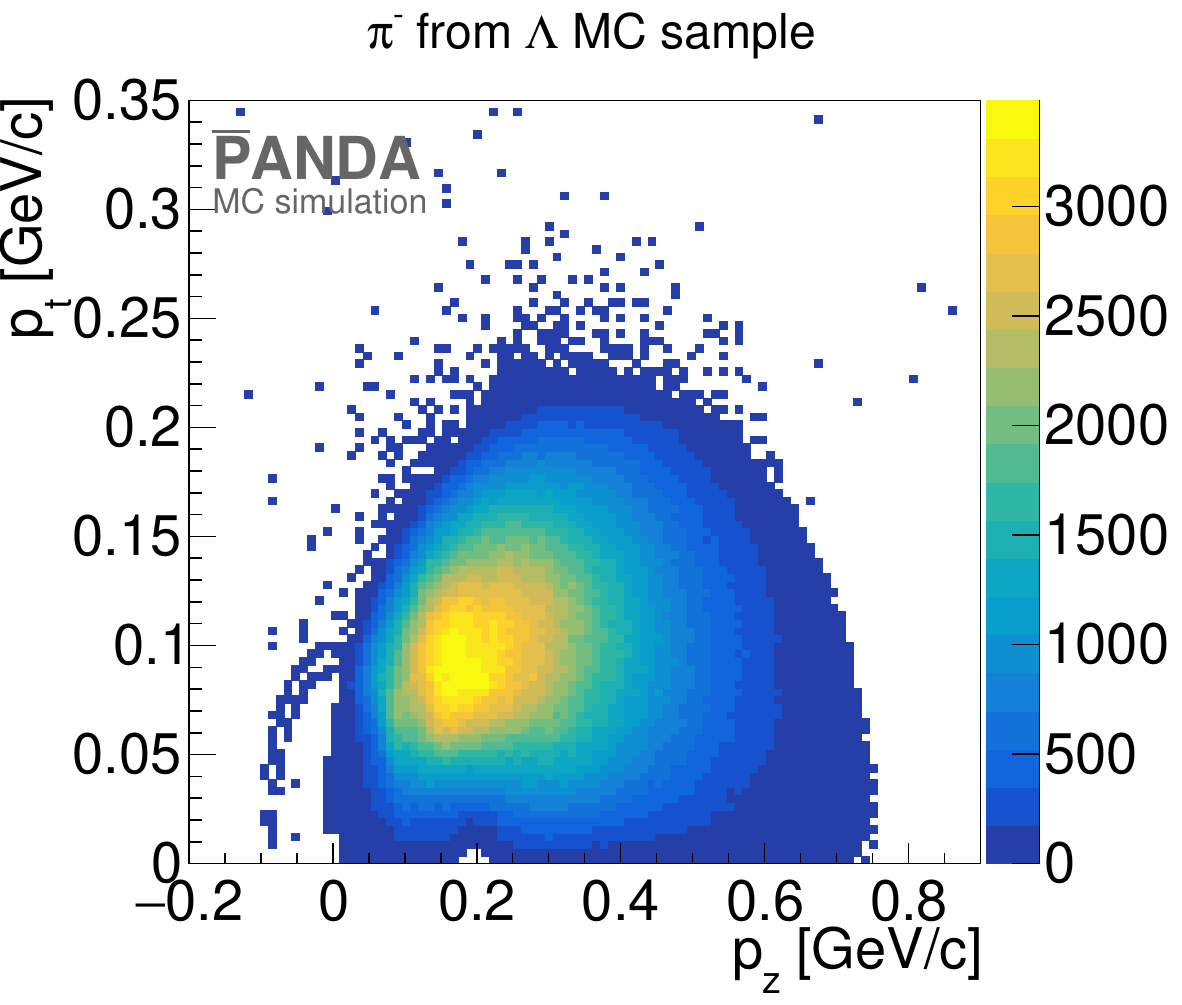}
			\caption{}
		\end{subfigure}
		\begin{subfigure}[b]{0.49\textwidth}
			\centering
			\includegraphics[width=0.8\textwidth]{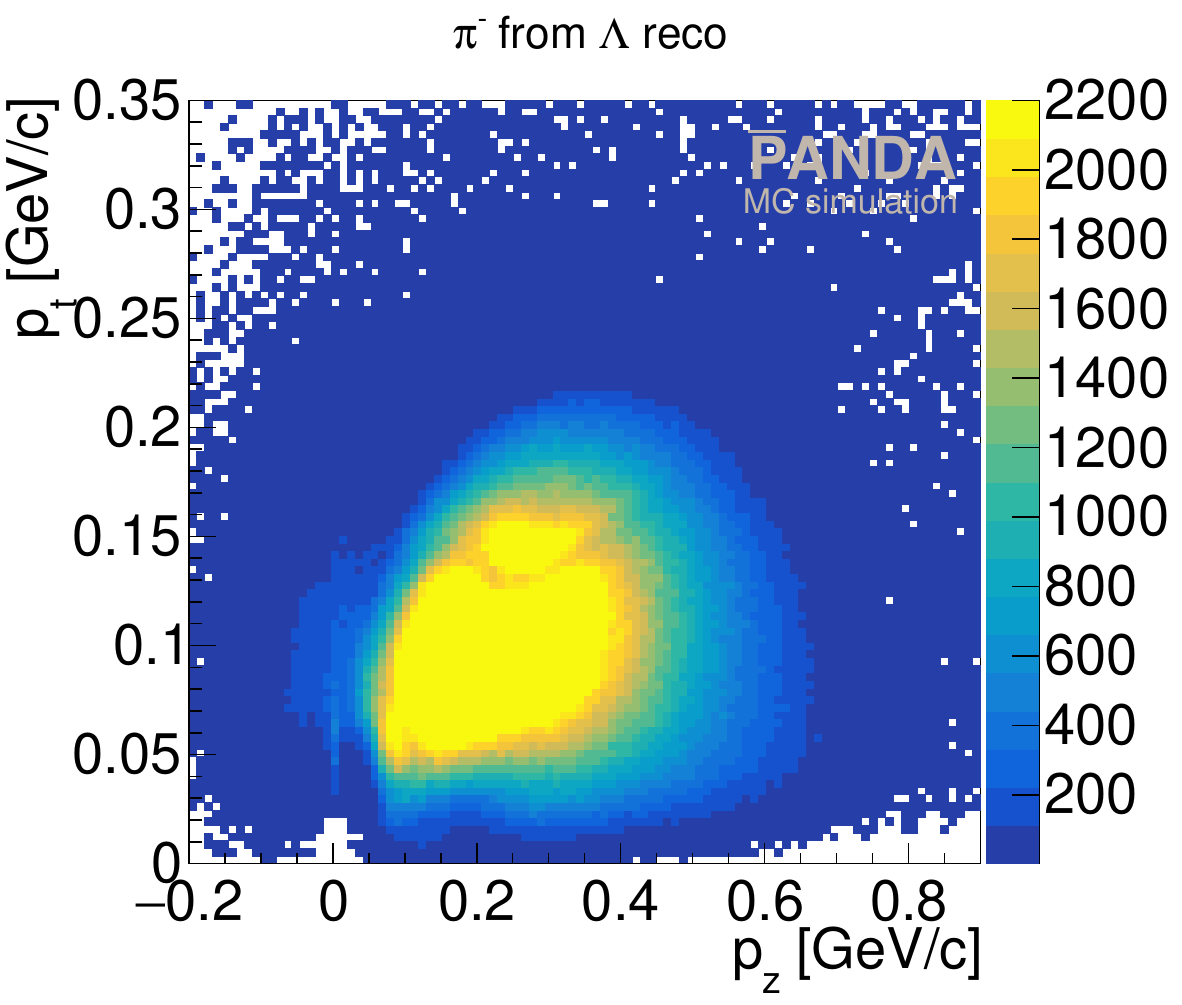}
			\caption{}
			\label{fig:PiMinusLamReco}
		\end{subfigure}
		\caption{Transverse vs. longitudinal momentum distribution for generated (a) and reconstructed (b) \piminus candidates from \lam, requiring that the generated \lam has only two daughters.}
		\label{fig:PiMinusLam}
	\end{figure}

\begin{figure}[htbp]
		\centering
		\includegraphics[width=0.9\linewidth]{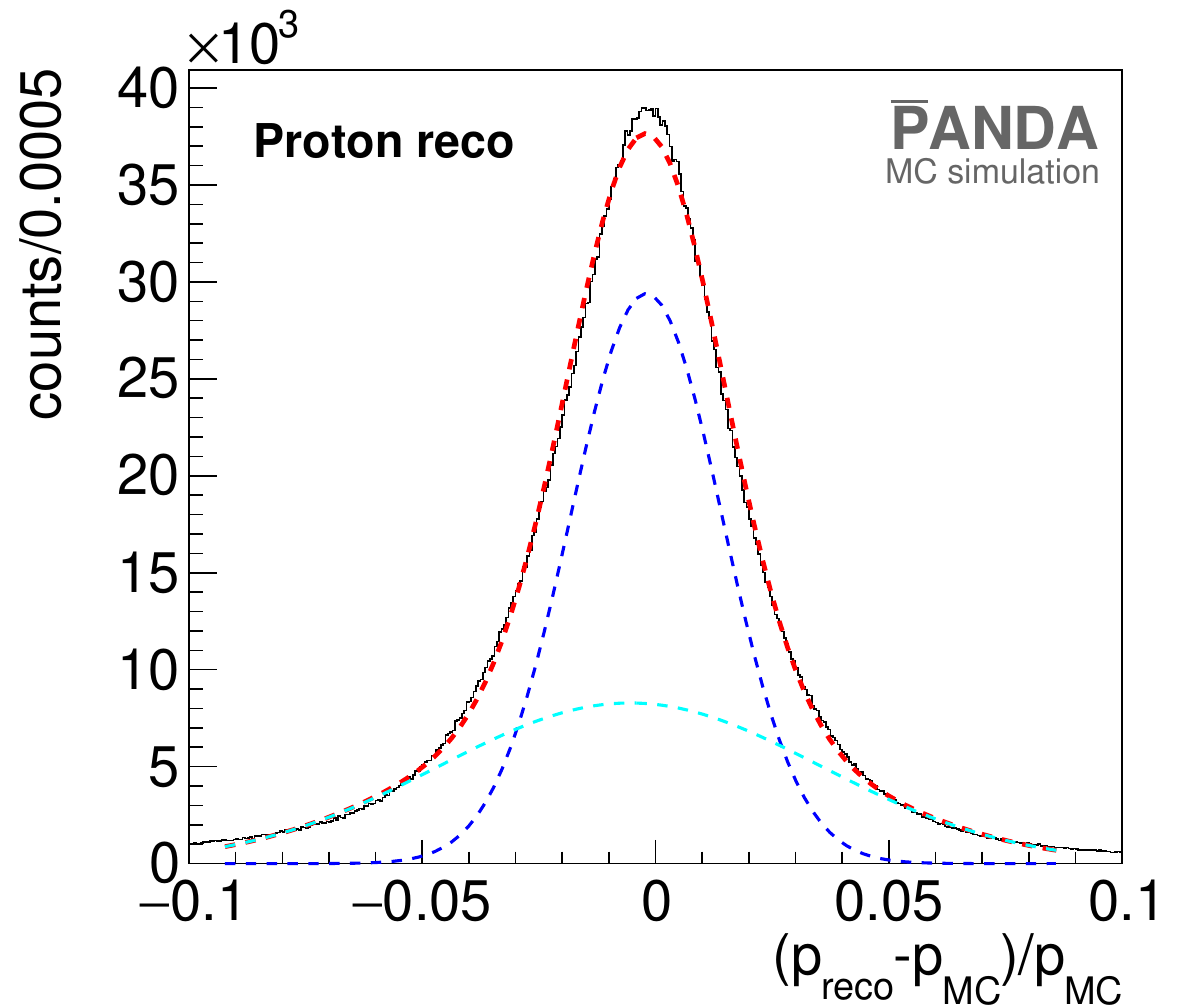}
		\caption{Relative momentum resolution of the reconstructed protons (black histogram). The distribution is fitted with a double Gauss fit (red dashed line). The width of the inner Gauss function (blue dashed line) is used as momentum resolution. The cyan dashed line indicates the second Gaus function.}
		\label{fig:DoubleGaussFitProton}
	\end{figure}

	After the track filtering, the final state particle candidates are filled into the corresponding candidate lists.
	In the most pessimistic scenario, no information about the particle species is provided by the detector.
	Therefore, no PID information is used for the selection of the possible candidates, meaning that only the information about their charge is used.
	For a given charge sign each of the corresponding candidate lists is filled with the same candidate.
	This implies that a positive charged particle is filled into the proton, \piplus and \kplus list, while a particle with negative charge is filled into the \aprot, \piminus and \kminus list, respectively.
	The single candidates differ only in the mass, which is set according to the hypothesis of the corresponding candidate list.\\
	If at least three candidates for each charge sign are available per event, it is marked as "reconstructable".
	This pre-selection avoids the reconstruction of incomplete signal events.\\
	In the following, the reconstruction efficiency is defined as the ratio of MC matched candidates to the number of generated candidates.
	MC matched means that the reconstructed candidate has a partner in the MC truth list which has the correct event genealogy up to the initial \pbarp system.	
	\begin{table}[b]
		\centering
		\caption{Reconstruction efficiency for the final state particles of \mychannelfs and \mychannelfscc (c.c.), respectively.}
		\label{tab:RecoEffFinalStates}
		\begin{tabular}{lrr}
			\hline
			particle type & eff. [\%] & eff.[\%](c.c.)\\
			\hline
			\piminus & 71.2 & 70.6\\
			\piplus (\alam) & 68.6 & 68.3 \\
			\piplus (\anticascade) & 73.7 &  73.1\\
			\kminus (resonance) & 84.9 & 86.7\\
			\kminus (continuum) & 85.1 & 86.9\\
			p & 88.7 & 86.2\\
			\aprot & 82.3 & 83.4\\
			\hline
		\end{tabular}
	\end{table}
	The reconstruction efficiencies achieved for the final state particles are listed in \cref{tab:RecoEffFinalStates}.
	The statistical error on the reconstruction efficiency is of the order of $0.1\,\%$.
	A systematic error, for example caused by the acceptance of the individual sub-detectors, is not included.
	Since the reconstruction efficiency of the final state particles depends on their production point,  the efficiencies differ for the different particles.\\
	For each final state particle two-dimensional histograms of transverse momentum versus longitudinal momentum as well as absolute momentum versus polar angle are generated.
	As an example, the generated and the reconstructed transverse versus longitudinal momentum distributions for \piminus coming from \lam decay are shown in \cref{fig:PiMinusLam}.
	Here, the generated distributions are used as reference plots to deduce the quality of the reconstruction.
	The discontinuity observable in \cref{fig:PiMinusLamReco} is caused by the transition area between the Target Spectrometer and the Forward Spectrometer.
	For all final state candidates the distributions contain entries outside the kinematically allowed.
	This could be caused by interactions of the generated particles inside the detector material or with the beam pipe during the propagation.
	In addition, the generated distribution shows an ellipse of entries which corresponds stopped \lam that subsequentially decay into a p\piminus pair.
	The comparison between the generated and the reconstructed distributions shows that the \piminus from the signal events are clearly identifiable.\\
	The relative momentum resolution is obtained from
	\begin{equation}
		\frac{\Delta p}{p}=\frac{p^{\mt{reco}} - p^{\mt{MC}}}{p^{\mt{MC}}}
	\end{equation}
	where $p^{\mt{reco}}$ denotes the reconstructed and $p^{\mt{MC}}$ the generated momentum.
	The value of the resolution is determined by performing a double Gaussian fit to the resulting distribution, see \cref{fig:DoubleGaussFitProton}.
	The width of the inner, most narrow, Gauss function is used as the momentum resolution.
	Here, about $64\,\%$ of the yield is within the range of the inner Gauss function and about $86\,\%$ in the range of the second Gauss function.
	By varying the fit parameters, the systematic error of the fit value is estimated to be $0.09$ percentage points.
	\begin{table}[t]
		\centering
		\caption{Momentum resolution for the final state particles of \mychannelfs and \mychannelfscc (c.c.), respectively. The error on the fit value is dominated by the systematic error which is estimated to be $0.09$ percentage points.}
		\label{tab:MomResFinalStates}
		\begin{tabular}{lrr}
			\hline
			particle type & $\Delta p/p$ [\%] & $\Delta p/p$ [\%](c.c.)\\
			\hline
			\piminus & 1.61 & 1.61\\
			\piplus (\alam) & 1.64 & 1.64\\
			\piplus (\anticascade) & 1.48 & 1.48 \\
			\kminus (res.) & 1.65 & 1.65 \\
			\kminus (cont.) & 1.66 & 1.65 \\
			p & 1.63 & 1.61 \\
			\aprot & 1.59  & 1.60 \\
			\hline
		\end{tabular}
	\end{table}
	The determined fit values are summarized in \cref{tab:MomResFinalStates}.
	\subsubsection*{\textbf{Intermediate State Particles}}
	\begin{figure*}
		\begin{subfigure}[b]{0.49\textwidth}
			\includegraphics[width=0.8\textwidth]{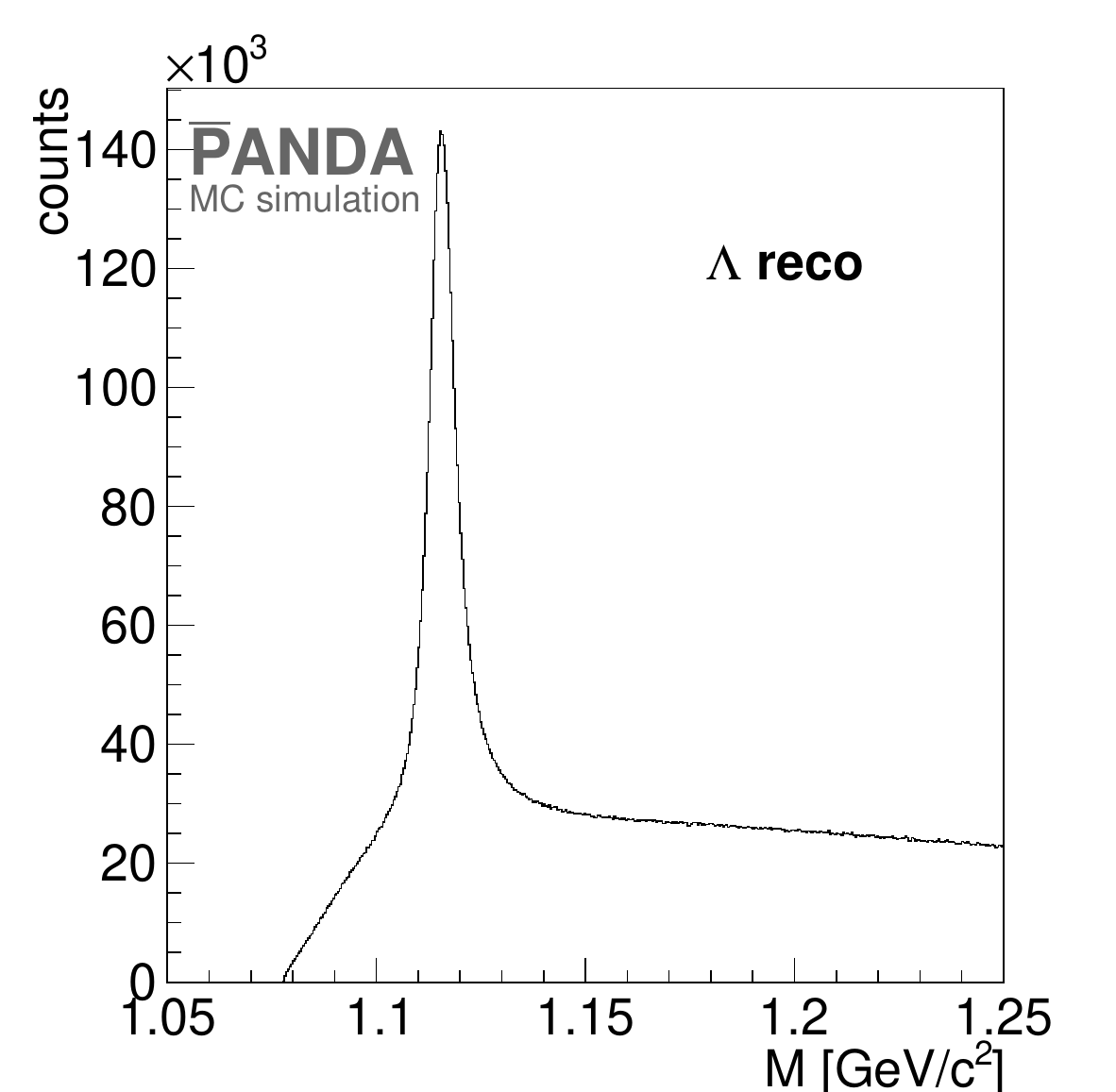}
			\caption{}
		\end{subfigure}
		\begin{subfigure}[b]{0.49\textwidth}
			\includegraphics[width=0.8\textwidth]{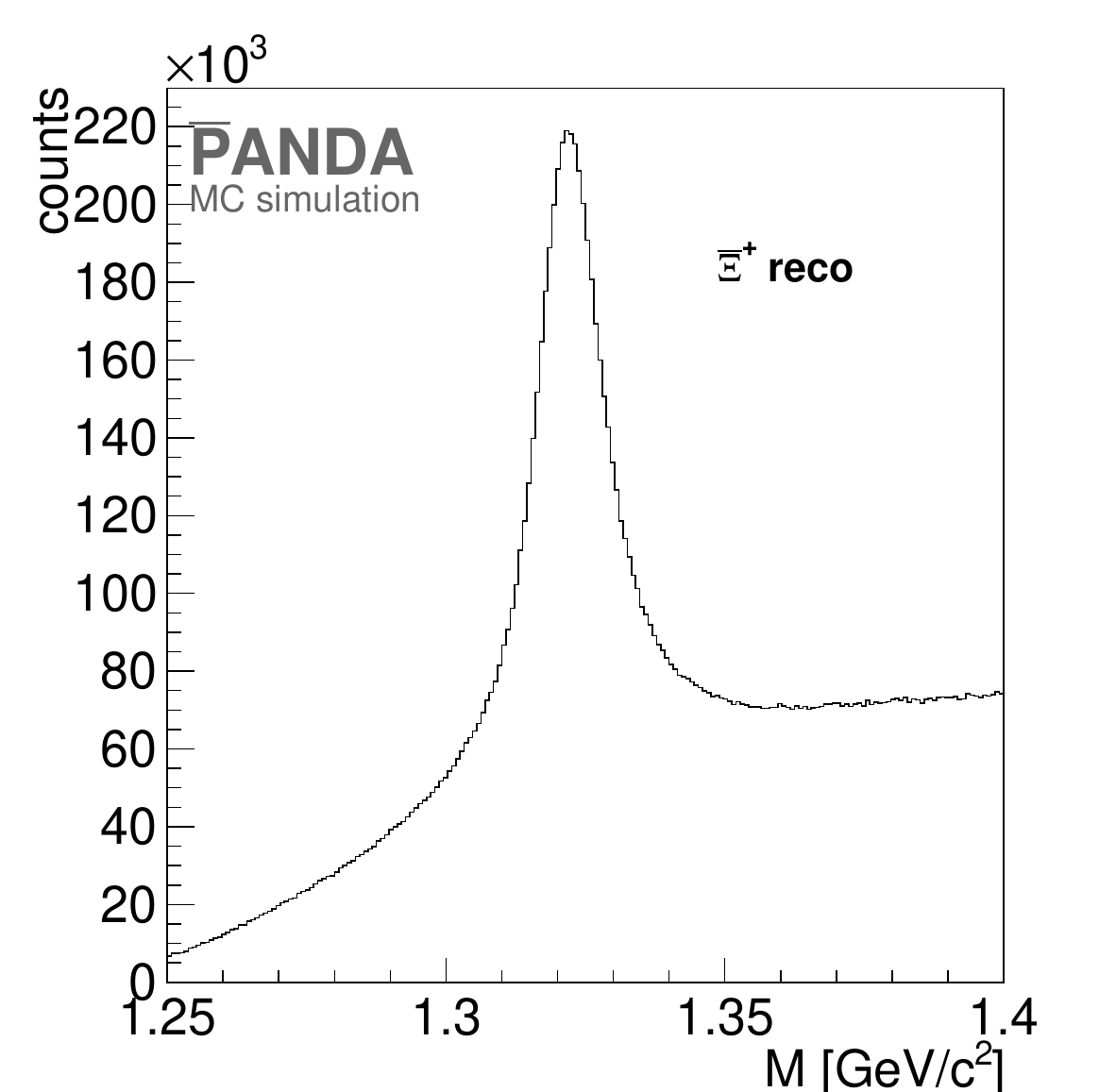}
			\caption{}
		\end{subfigure}
		\caption{Invariant mass spectrum of \lam (a) and \anticascade (b) after the mass window selection. \lam and \anticascade are produced in the reaction \mychannelfs.}
		\label{fig:HyperonMassSpectraMassWindowCut}
	\end{figure*}
	The candidate selection of the intermediate state particles, i.e. \alam, \lam, \anticascade and \cascade, are similar for each particle type.
	In the first step, \alam and \alam are built by combining the daughter particles: \aprot and \piplus for \alam, p and \piminus for \lam.
	In the next stage of reconstruction \alam and an additional \piplus are combined to \anticascade as well as \lam and \piminus to \cascade in the \cc channel.
	Since the input for the \DTF are \enquote{raw} candidates, only a coarse pre-selection is done to reduce the number of wrongly combined candidates.
	For this a symmetric mass window selection of $\pm 0.15$\massunit around the nominal hyperon mass is applied on the candidate masses.
	The \lam and \anticascade invariant mass spectra after the mass window selection are shown in \cref{fig:HyperonMassSpectraMassWindowCut}.
	This selection rejects candidates with a mass much higher than the input hyperon mass.
	All remaining candidates are passed to the next stage of reconstruction.
	%
	\subsubsection*{\textbf{Full Decay Tree}}
\begin{figure}[h]
		\centering
		\includegraphics[width=0.4\textwidth]{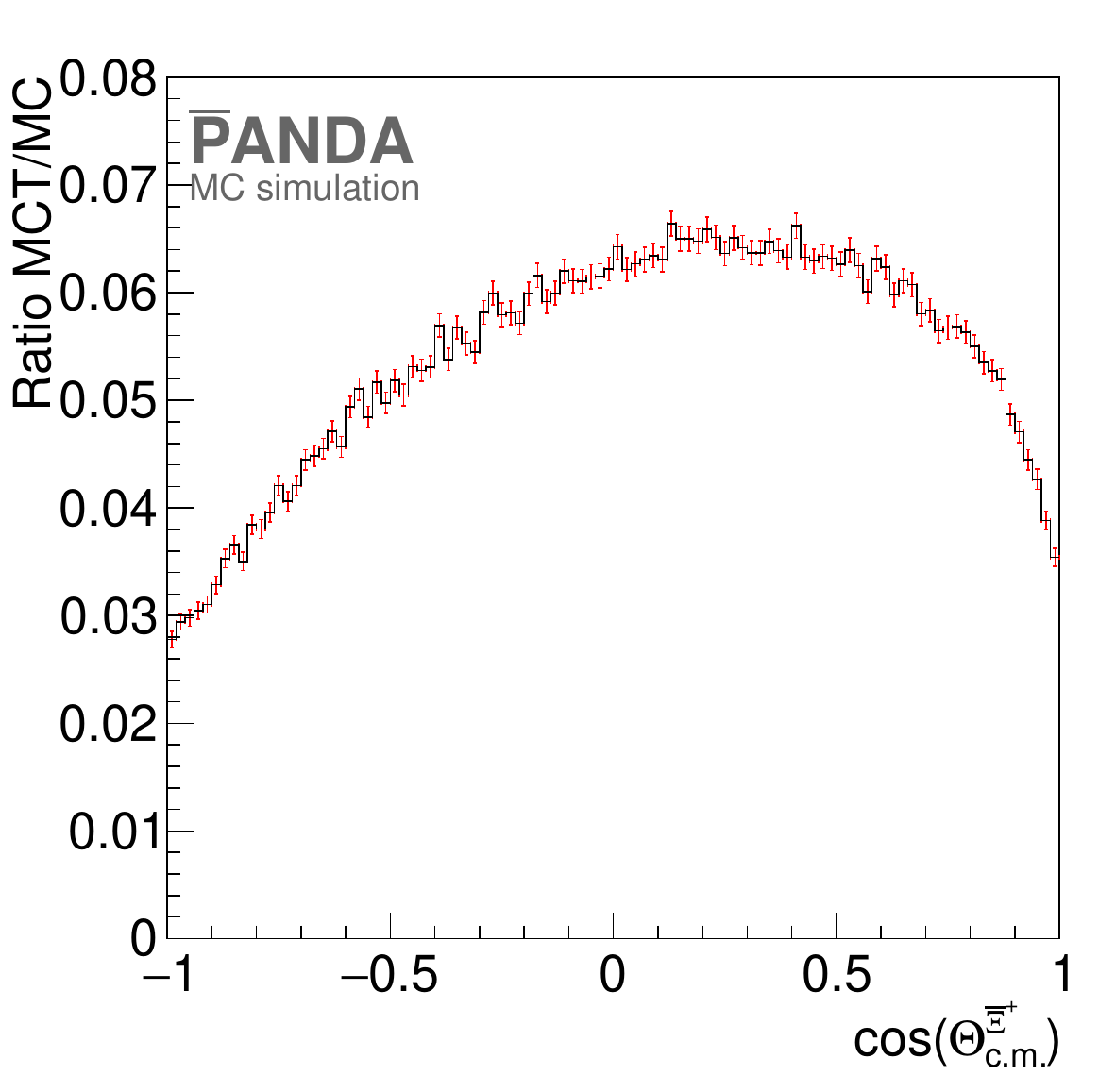}
		\caption{Ratio of the $\cos\left(\Theta\right)$ distributions in the c.m. frame for the final selected \anticascade candidates in the process \mychannelfs. MC indicates the generated candidates and MCT the MC truth partners of the final selected candidates.}
		\label{fig:CosThtRatio}
	\end{figure}

	In the following, the reconstruction of the full decay tree is described.
	Within this procedure, described in \cref{sec:DecayTreeFit}, the four-momentum conservation of the initial energy and momentum vector 
	\[
	P_{\mt{ini}} = \left(0,0,4.6,5.633\right)\unit{GeV},
	\]
	as well as the hyperon masses are constraint.
	Unless otherwise indicated, the results listed below are for the \fs final state.\\
	Since the $\Xi$ resonances decay promptly into a \lam \kminus pair or into \alam \kplus in the \cc channel, the reconstruction of the full decay tree is done by combining \fs and \fscc, respectively.
	Subsequently, the candidates are fitted with the \DTF implemented in PandaRoot.
	The fit quality is represented by the \chisq value and a fit probability is calculated.
	\begin{figure*}[htb]
		\begin{subfigure}[b]{0.49\textwidth}
			\centering
			\includegraphics[width=0.8\textwidth]{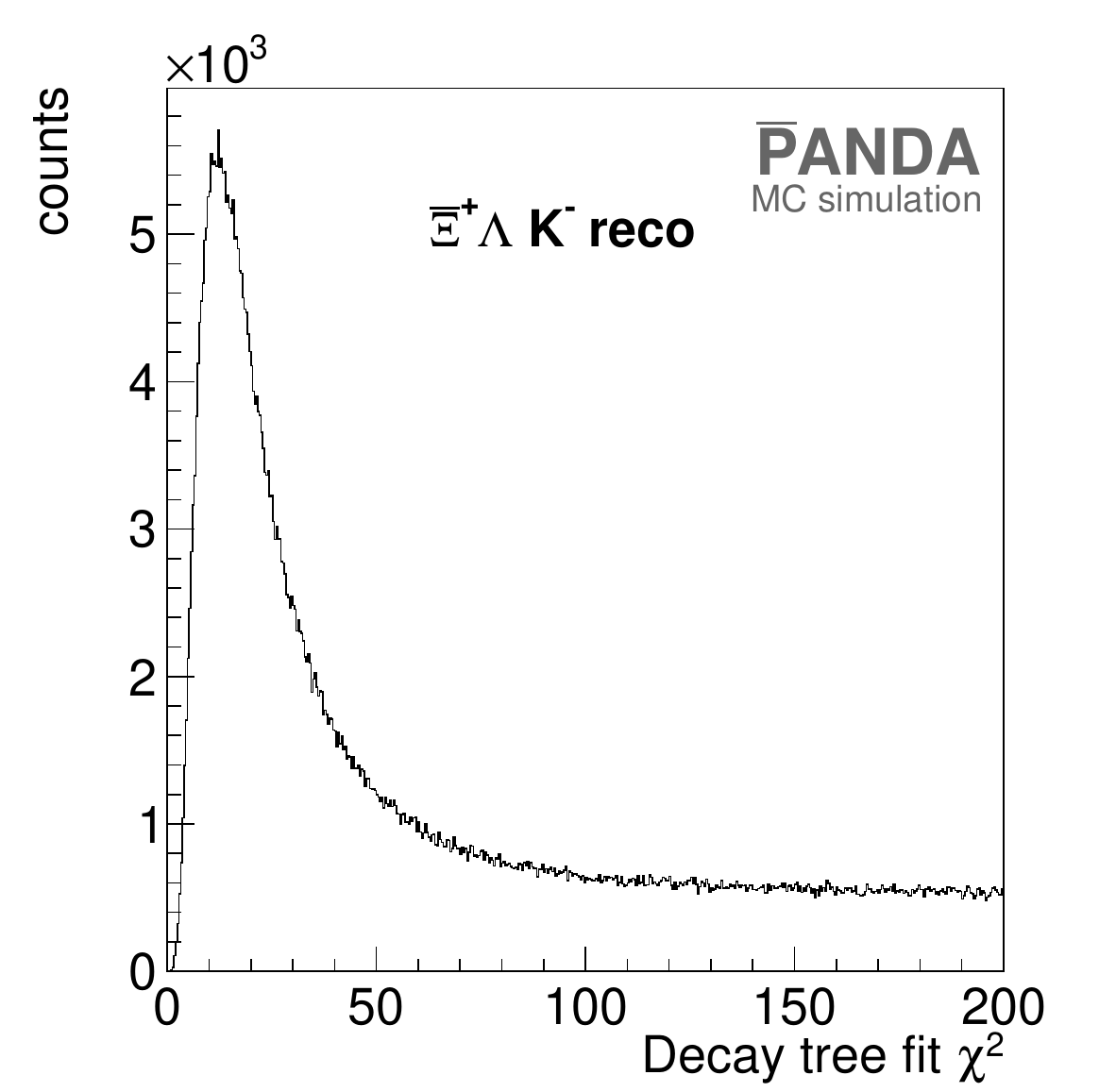}
			\caption{}
		\end{subfigure}
		\begin{subfigure}[b]{0.49\textwidth}
			\centering
			\includegraphics[width=0.8\textwidth]{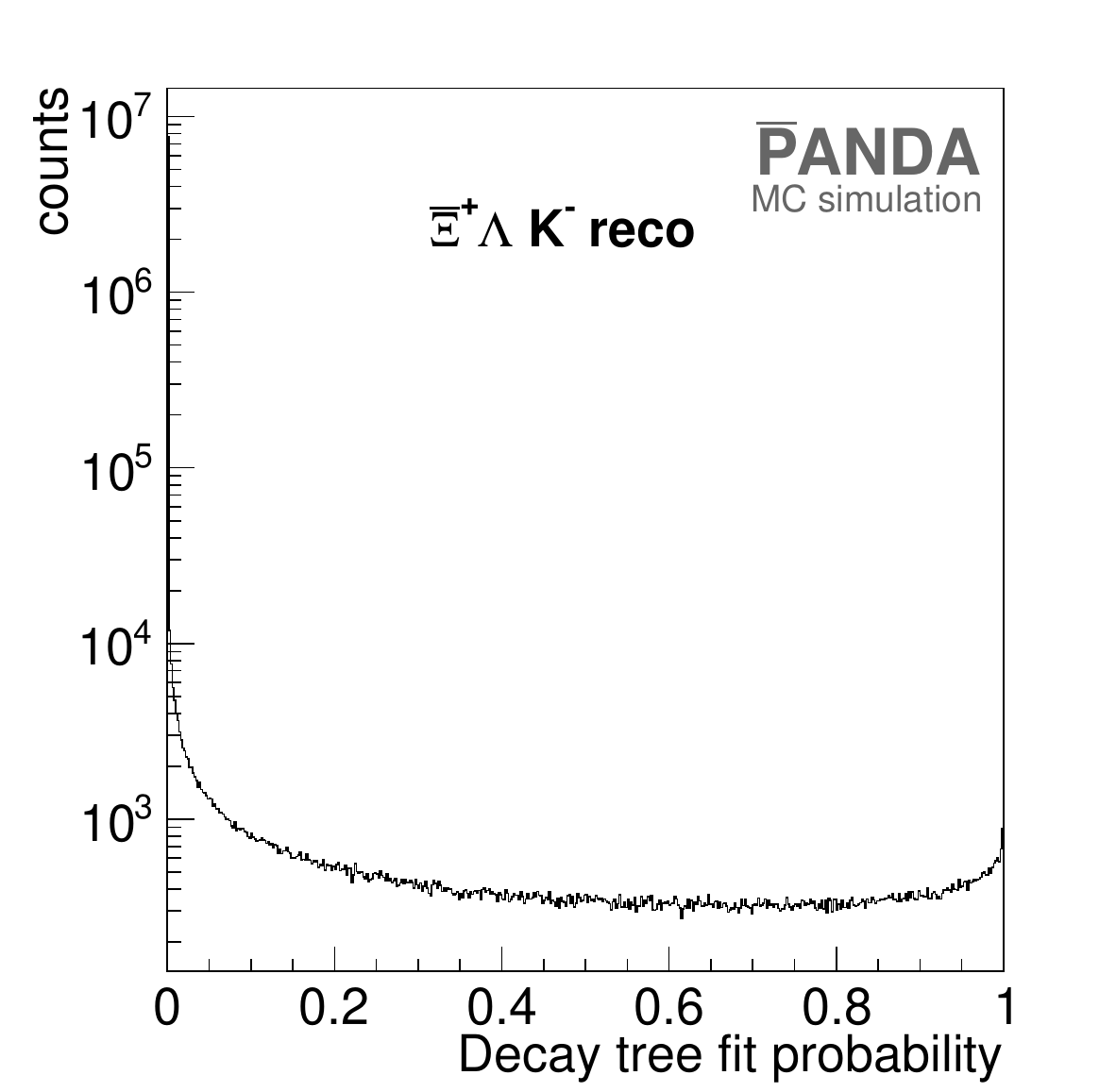}
			\caption{}
			\label{fig:DTF_prob}
		\end{subfigure}
		\caption{\chisq (a) and probability (b) distribution for the decay tree fit performed on the \fs sample. The rise of the probability distribution indicates that the errors are overestimated in some cases.}
		\label{fig:DTFQuality}
	\end{figure*}
	\Cref{fig:DTFQuality} shows the corresponding distributions.
	The probability distribution (Fig. \ref{fig:DTF_prob}) shows a rising behaviour close to the value of one.
	That indicates that the errors are overestimated for some cases.
	For the final selection only candidates, which have been successfully fitted are taken into account.
	The fit probability (P) is used as selection criterion for the candidate selection.
	Here, a lower threshold of $P>1\cdot 10^{-4}$ is applied corresponding to a selection on the \chisq value with $\chi^{2}<43$.
	The applied selection criterion was optimized according to reach the best figure of merit in terms of reconstruction efficiency and pure signal fraction of the final selected sample.
	The final selected sample contains 277,133 \fscc events and 283,617 \fscc events.
	\Cref{tab:RecoEffFinal} summarizes the achieved reconstruction efficiency and the signal purity for the final selected signal samples.
	The achieve reconstruction efficiencies are strongly depending on the track efficiency of the tracking algorithms which is by default $100\,\%$ for the ideal pattern recognition algorithm.
	By assuming different track efficiencies, i.e. $90\,\%$, $85\,\%$, and $80\,\%$ the signal reconstruction efficiency for the \fs sample is reduced to $3\,\%$, $2.9\,\%$, and $2.1\,\%$, respectively.\\	
	In addition to the reconstruction efficiency, the ratio between the resonant and 
	\begin{table}[b]
		\centering
		\caption{Reconstruction efficiency and purity for the final selected signal samples.}
		\label{tab:RecoEffFinal}
		\begin{tabular}{lrr}
			\hline
			Sample & Reco. Eff. [\%] & Purity [\%] \\
			\hline
			\fs & 5.4 & 97.7 \\
			\fscc & 5.5 & 97.7 \\
			\hline
		\end{tabular}
	\end{table}
	the non-resonant decay modes is determined, see \cref{tab:RatioDecayModes}.
	\begin{table}[b]
		\centering
		\caption{Channels and their fraction of ther generated cross section for the \mychannelfs and \mychannelfscc final reconstructed sample (Reco.) and the generated sample (Input).}
		\label{tab:RatioDecayModes}
		\begin{tabular}{lrr}
			\hline
			Channel & Reco. [\%] &  Input [\%]\\
			\hline
			\anticascade \excitedcascadesixteen & $37.7 \pm 0.8$ & 40 \\
			\anticascade \excitedcascadetwenty & $42.4 \pm 0.8$ & 40 \\
			\fs & $19.9 \pm 0.5$ & 20 \\
			\hline
			\vspace{.5pt}\\
			\cascade \excitedanticascadesixteen & $37.8 \pm 0.8$ & 40\\
			\cascade \excitedanticascadetwenty & $42.2 \pm 0.8$ & 40\vspace{2pt}\\
			\fscc & $19.9 \pm 0.5$ & 20 \\
			\hline
		\end{tabular}	
	\end{table}
	Due to different efficiencies for the excited $\Xi$ states, the determined fraction differs from the input values.
	The fraction for the continuum contribution is in good agreement with the input.
	\begin{figure}[t]
		\centering
		\includegraphics[width=0.49\textwidth]{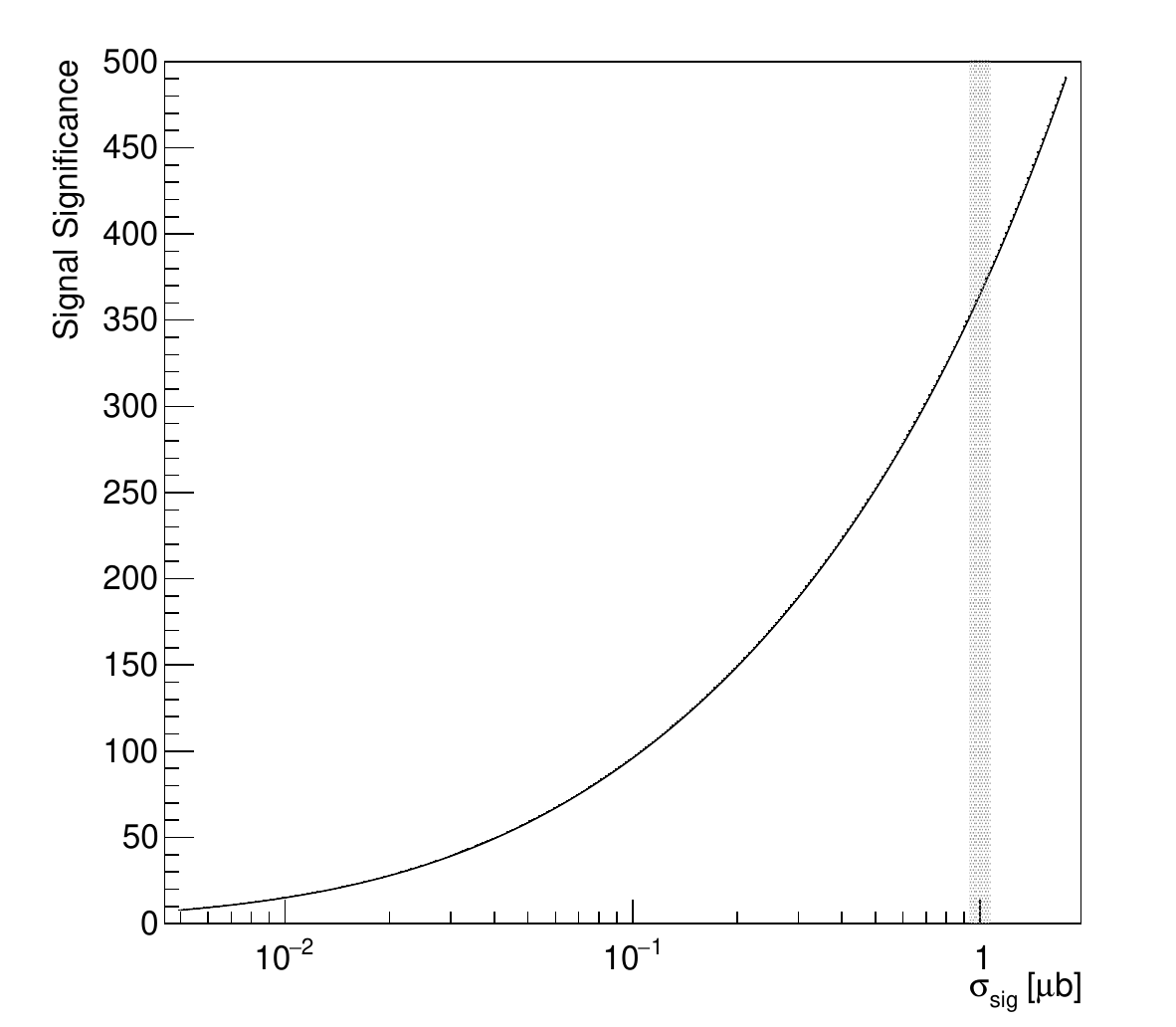}
		\caption{Signal significance as function of the signal cross section. The gray band indicates the region the signal is assumed.}
		\label{fig:SigvsXSec}
	\end{figure}
	The signal significance is defined as $S_{\mt{Sig}}=S/\sqrt{S+B}$ and depends on the cross section of the reaction to study, see \cref{sec:BackgroundStudies}, called signal cross section in the following.
	Here, $S$ and $B$ are the number of signal and background events, respectively.
	\Cref{fig:SigvsXSec} shows the expected signal significance as function of the signal cross section.
	The signal final state is clearly identifiable above the hadronic background, even if the cross section is an order of magnitude smaller than assumed here.\\
	The \DTF uses the four-momentum constraint fit, which leads to a correction of the momentum and the energy for each involved candidate to match the initial four-momentum vector.
	This correction has an impact on the momentum resolution of the candidates.
	The momentum resolution is evaluated by performing a double Gaussian fit to the relative deviation of the reconstructed and generated total momentum, like described in \cref{sssec:FinalStateSelection}.
	\Cref{tab:MomResFinal} summarizes the evaluated momentum resolution of the intermediate state particles.\\
	From the deviation of the reconstructed from the generated decay vertex position of all three spatial coordinates the decay vertex resolution is determined.
	\begin{figure}
		\includegraphics[width=0.4\textwidth]{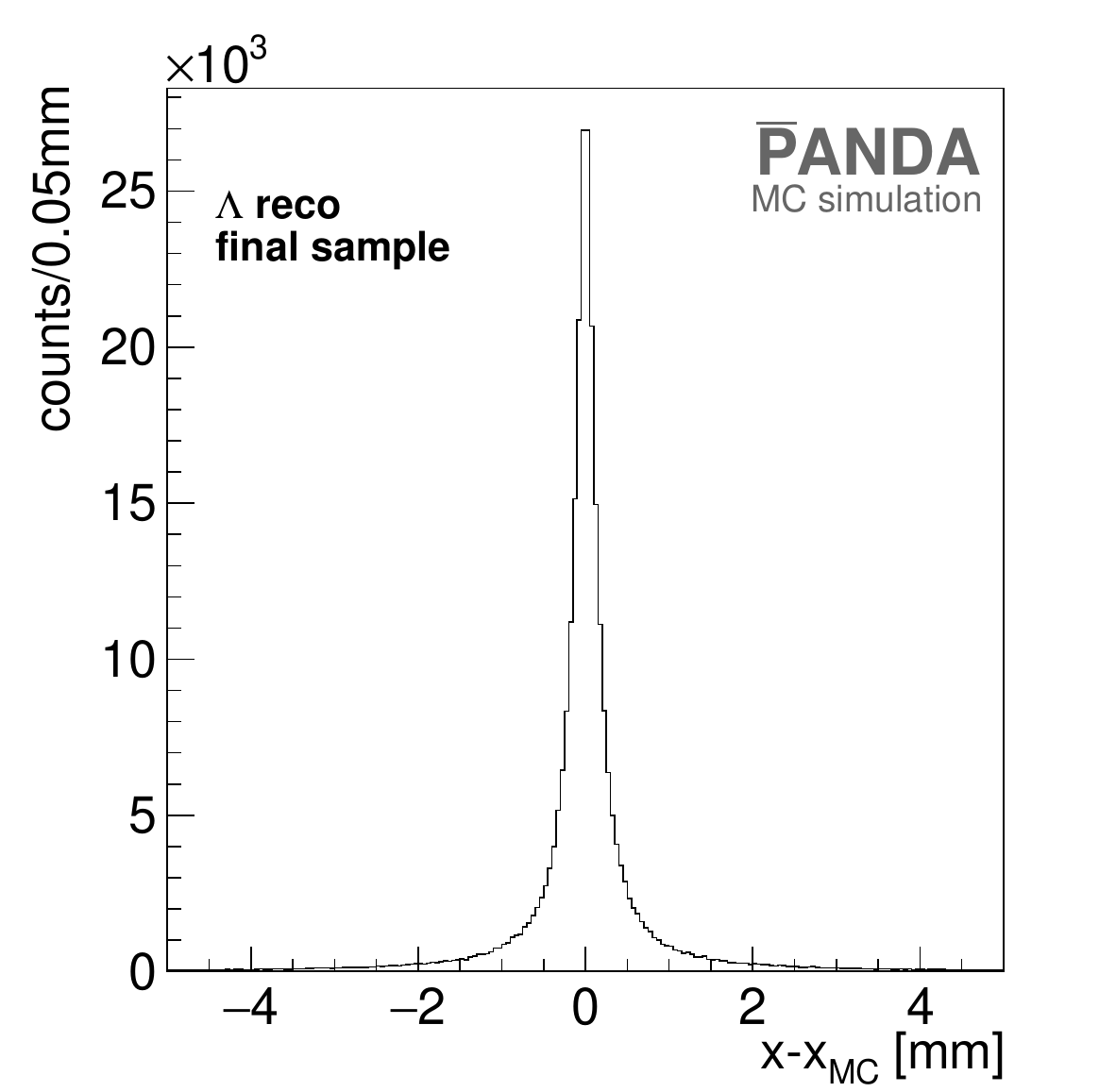}
		\caption{Deviation of the reconstructed from the generated x coordinate of the \lam decay vertex in the process \mychannelfs after the final selection.}
		\label{fig:VtxResFinal_Lam}
	\end{figure}
	\Cref{fig:VtxResFinal_Lam} shows the deviation of the decay vertex position of final selected \lam for the x coordinate as an example.
	The resulting distribution is clearly not Gaussian.
	Therefore, the decay vertex resolution is determined by evaluating 
	\begin{equation}
		\sigma_{\mt{vtx}} = \frac{\mt{FWHM}}{2\cdot \sqrt{2\cdot \ln 2}},
	\end{equation}
	where FWHM is the full width at half maximum of the distribution.
	The achieved resolutions for all intermediate state particles are listed in \cref{tab:VtxRes}.
	\begin{table}[b]
		\caption{Relative momentum resolution for the intermediate state particles of \mychannelfs and \mychannelfscc.}
		\label{tab:MomResFinal}
		\centering
		\begin{tabular}[b]{lc}
			\hline
			Particle & $\sigma_{\mt{p}}\,\left[\%\right]$ \\
			\hline
			\lam & $\left( 0.777\pm0.007\right)$\\
			\alam & $\left( 0.803 \pm 0.007\right)$ \\
			\anticascade & $\left( 1.30 \pm 0.01\right)$ \\
			\hline
			\vspace{0.1pt}\\
			\lam & $\left( 0.795 \pm 0.006\right)$ \\
			\alam & $\left( 0.748 \pm 0.006\right)$ \\
			\cascade & $\left( 1.29 \pm 0.01\right)$ \\
			\hline
		\end{tabular}
	\end{table}
	\begin{table}[h]
		\centering
		\caption{Decay vertex resolution for each spatial direction of the final selected intermediate state particles of \mychannelfs and \mychannelfscc.}
		\label{tab:VtxRes}
		\begin{tabular}[t]{lrrr}
			\hline
			Particle & x [mm] & y [mm] & z [mm] \\
			\hline
			\multicolumn{4}{c}{\mychannelfs}\\
			\lam & 0.110 & 	0.093& 0.544 \\
			\alam & 0.127& 0.110& 0.595 \\
			\anticascade & 0.119& 0.119& 0.510 \vspace{0.1pt}\\
			\hline				\multicolumn{4}{c}{\mychannelfscc}\\
			\lam & 0.127& 0.110& 0.578 \\
			\alam & 0.110&	0.110& 0.544 \\
			\cascade & 0.119& 0.119&	0.510 \\
			\hline
		\end{tabular}
	\end{table}
	Since the determined FWHM is depending on the chosen bin size, the error on the FWHM is estimated by varying the number of bins of the corresponding histogram.
	With this procedure, the error on the vertex resolution is estimated to be about $8\,\mu\mt{m}$.\\
	The decay products of the resonance together with the additional hyperon, \fs and \fscc, can be defined as a three-body final state of the strong interaction, since the involved particles further decay weakly or electromagnetically.
	In this analysis, $M^2$(\lam\kminus) and $M^2$(\anticascade\kminus) as well as the squared mass for their \cc particles are used as the axes of the corresponding Dalitz plot.
	The different decay modes of the reaction lead to different distributions within the Dalitz plot. 
	For the continuum production of the three-body final state, the Dalitz plot shows a uniform distribution over the entire kinematically allowed region.
	For a contributing resonant process, the resonance will 
	\begin{figure}[b]
		\centering
		\includegraphics[width=0.4\textwidth]{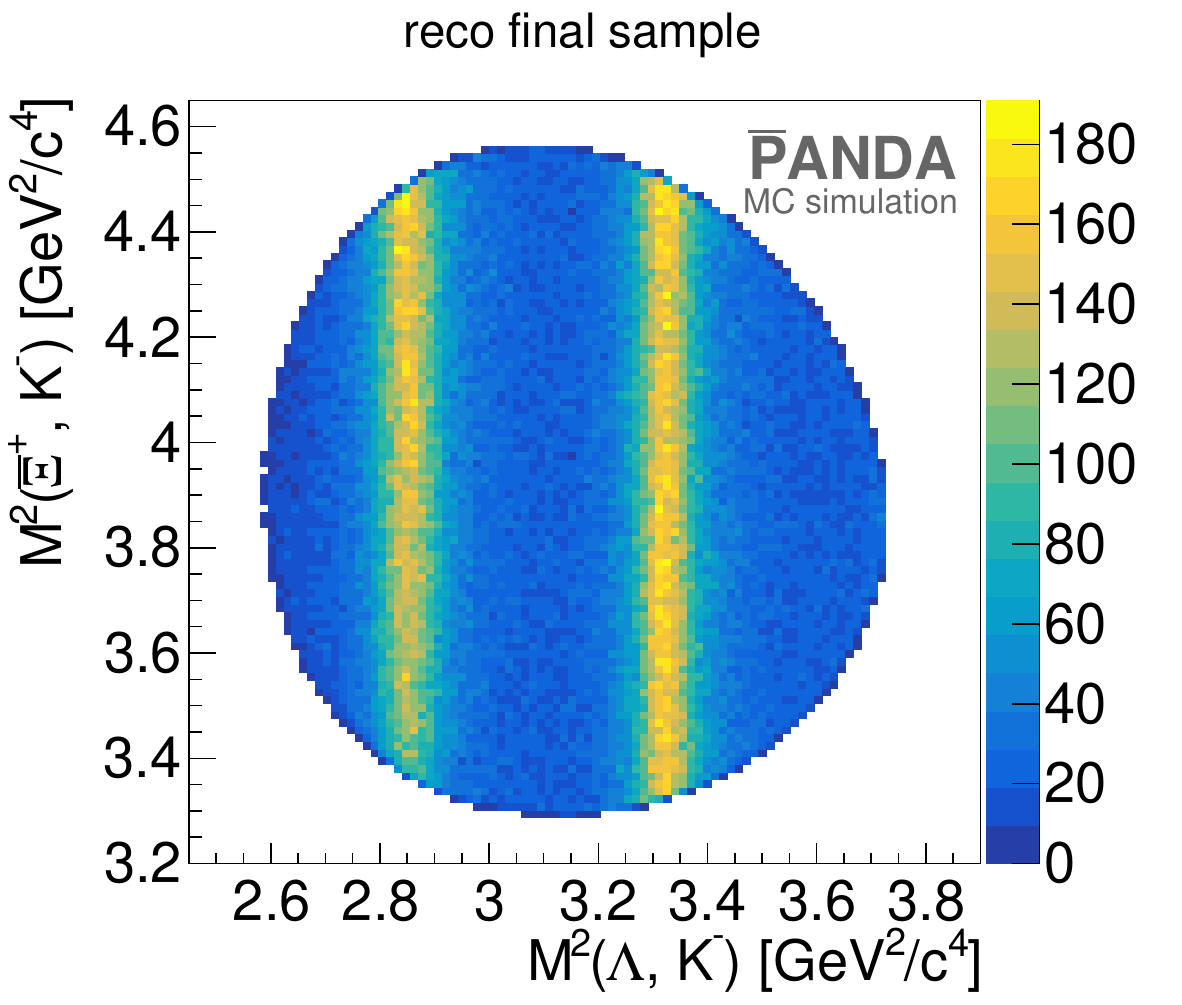}
		\caption{Dalitz plot for the final selected \fs candidates from \mychannelfs.}
		\label{fig:DalitzReco}
	\end{figure}
	be visible as structure in the Dalitz plot.
	The Dalitz plot for the \fs final state is shown in \cref{fig:DalitzReco}.
	Here, the $\Xi$ resonances are visible as vertical bands around the nominal squared mass values.
	\begin{figure}[t]
		\centering
		\includegraphics[width=0.4\textwidth]{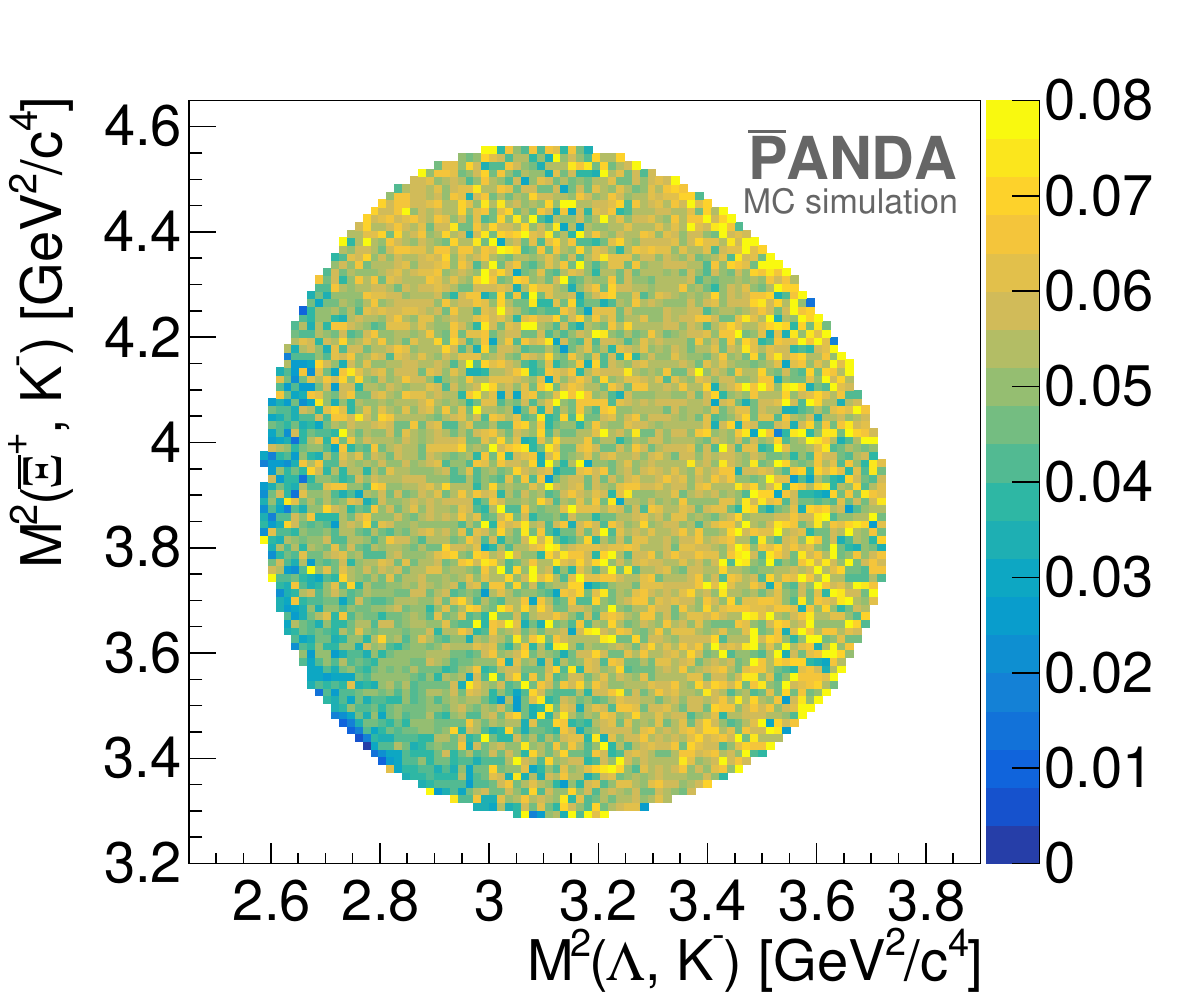}
		\caption{Ratio of the Dalitz Plots for the MC truth partners of the final \fs sample and the generated sample.}
		\label{fig:DalitzRatio}
	\end{figure}
	To compare the reconstructed and the generated Dalitz plot, the ratio of the Dalitz plots for the MC truth partners of the reconstructed and the generated candidates is illustrated in \cref{fig:DalitzRatio}.
	The ratio plot shows a uniform distribution.
	\begin{figure}[b]
		\centering
		\includegraphics[width=0.4\textwidth]{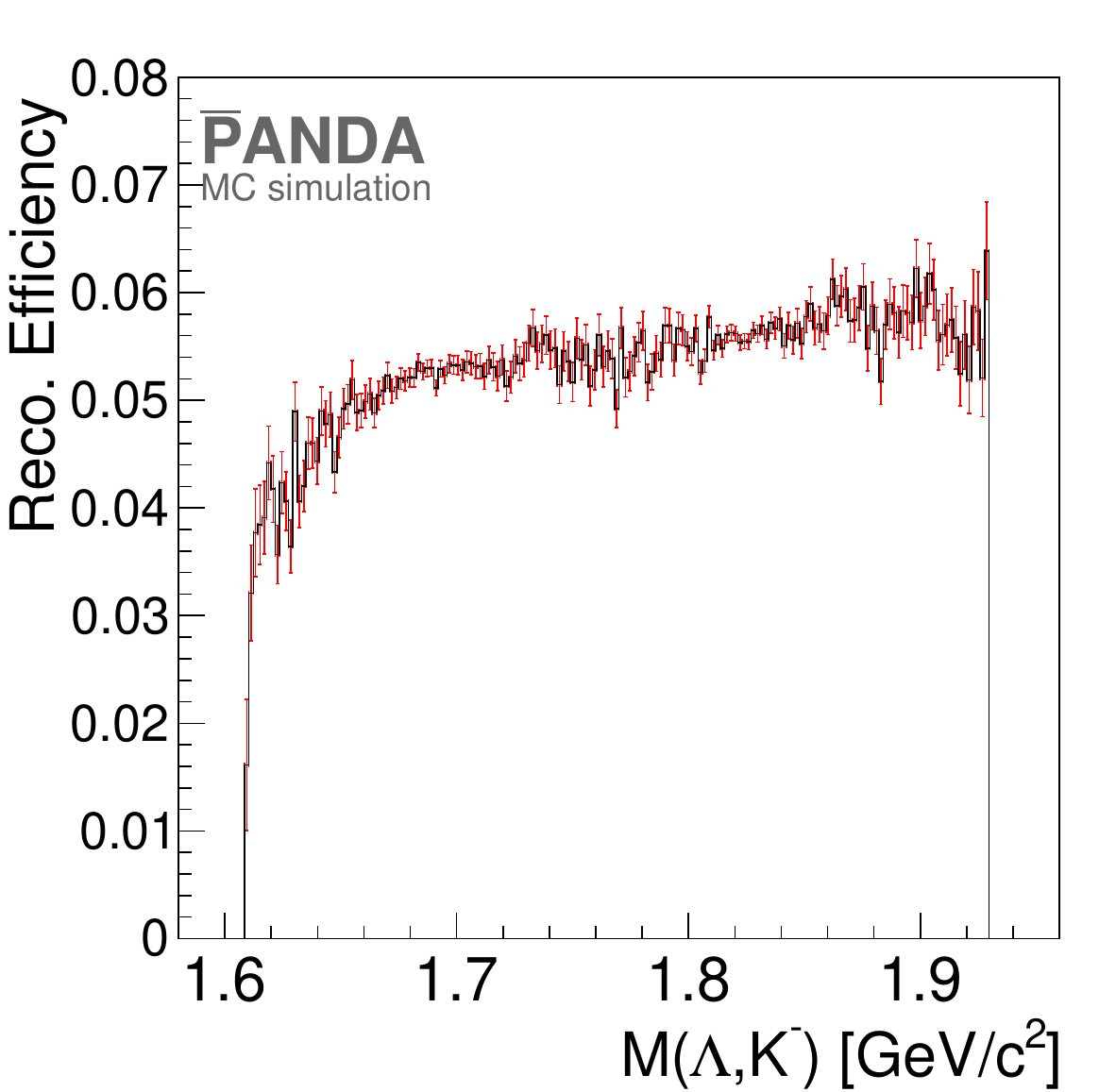}
		\caption{Reconstruction efficiency (black histogram) as function of the invariant \lam\kminus mass in the process \mychannelfs. The statistical error is shown in red.}
		\label{fig:LamKMassRatio}
	\end{figure}
	By illustrating the ratio of the generated and reconstructed mass distribution for the \lam\kminus sub-system, \cref{fig:LamKMassRatio}, one can observe a decrease of the reconstruction efficiency by about $20\,\%$ towards lower sub-system masses.\\
	The mass and the width of the resonances
	\begin{table}[h]
		\centering
		\caption{Fit results for the mass and width of the $\Xi$ resonances determined with a fit function containing two Voigt functions and a polynomial.}
		\label{tab:ResFitValues}
		\begin{tabular}{lcc}
			\hline
			& M [MeV$/\mt{c}^{2}$]  & $\Gamma$ [MeV$/\mt{c}^{2}$]  \\
			\hline
			\excitedcascadesixteen & $1689.99\pm 0.13$ & $30.1\pm 0.6$\\
			\excitedanticascadesixteen & $1690.16\pm 0.12$ &  $30.2\pm 0.6$ \\
			\excitedcascadetwenty & $1822.98\pm 0.12$ &  $22.9\pm 0.4$ \\
			\excitedanticascadetwenty & $1823.12\pm 0.12$ &  $22.7\pm 0.4$ \\
			\hline
		\end{tabular}
	\end{table}
	are determined by fitting a function containing two Voigt functions \cite{armstrong1967spectrum} and a polynomial to the corresponding mass distributions.
	\begin{figure}[b]
		\centering
		\includegraphics[width=0.4\textwidth]{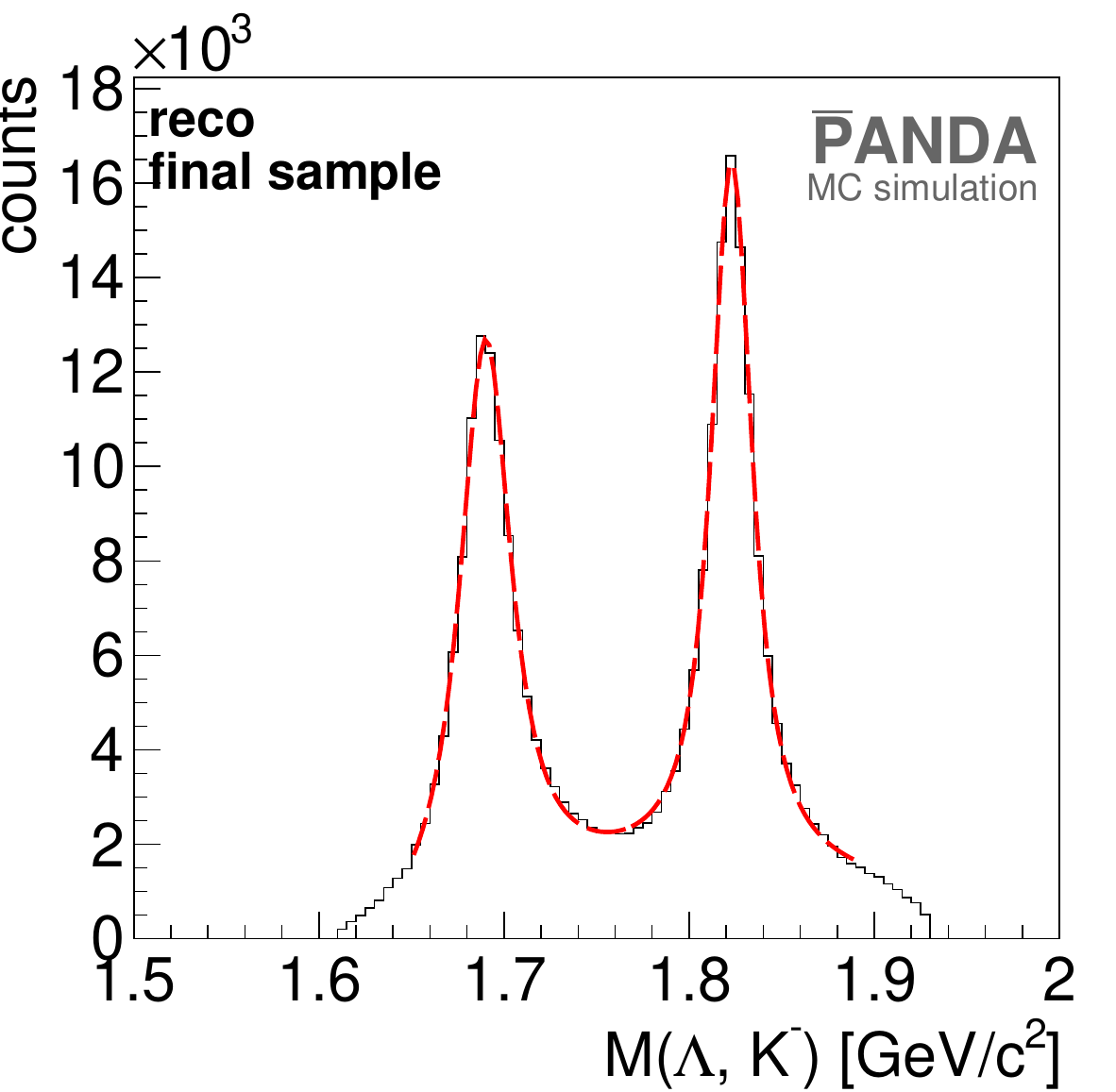}
		\caption{Mass distribution (black histogram) of the final reconstructed \lam\kminus from \mychannelfs with fit function (red dashed curve) containing two Voigt functions and a polynomial.}
		\label{fig:LamKMass}
	\end{figure}
	The mass distribution of \lam\kminus is shown as an example in \cref{fig:LamKMass}.
	In this analysis, the best fit result is achieved by fixing the instrumental width $\sigma_{M}$ for both resonances to $\sigma_{M}=4\unit{MeV}/c^2$.
	This value was determined by calculating the FWHM for the deviation of the final reconstructed and the generated mass distribution.
	The resulting fit values for the $\Xi$ resonances are summarized in \cref{tab:ResFitValues}.
	Except for the width for \excitedcascadetwenty and \excitedanticascadetwenty, the fitted values are consistent with the input values listed in \cref{tab:XiResProps}.
	The width for \excitedcascadetwenty and \excitedanticascadetwenty agree within $2\,\sigma$.\\
	\newline
	An isotropic angular distribution was assumed for the production of the \anticascade and \excitedcascade as well as for their \cc particles.
	From the ratio of the $\cos{\theta}$ distribution in the c.m. frame for the MC truth partners of the final selected candidates (MCT) and the generated (MC) candidates, shown in \cref{fig:CosThtRatio}, it is possible to deduce the reconstruction efficiency for any c.m. angular distribution.
	As it is indicated in \cref{fig:CosThtRatio}, the reconstruction efficiency will vary between $3\,\%$ and $6\,\%$ depending on the assumption. 
	Assuming a $\Delta$ function at $\cos{\theta}=0$, the reconstruction efficiency will be $6\,\%$ while the isotropic angular distribution gives about $5\,\%$.
	The ratio shows a reduced efficiency for particles emitted in forward and backward direction, which is due to the loss of propagated particles inside the beam pipe.
	%
	
	\subsection[$\bar{\mt{p}}\mt{p}\rightarrow \bar{\Xi^+} \Xi^- \pi^0$]{$\mathbf{\bar{\textbf{p}}\textbf{p}\rightarrow \bar{\Xi^+} \Xi^- \pi^0}$}
	\begin{figure}[htb]
		\centering
		\includegraphics[width=0.4\textwidth]{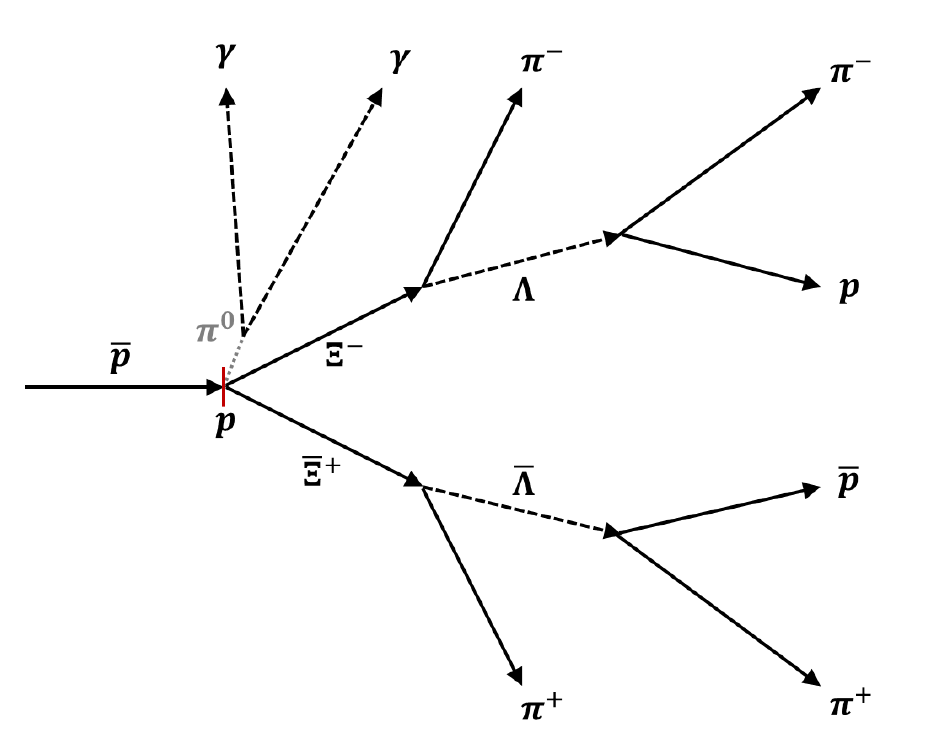}
		\caption{Schematic illustration of the decay tree for the process \pbarp$\rightarrow$ \cascasbarpinull.}
		\label{fig:DecayTreeAlbrecht}
	\end{figure}
	9 million signal events, generated according to the decay tree shown in \cref{fig:DecayTreeAlbrecht} have been analyzed containing a continuum contribution as well as the resonant states \excitedcascadefifteen, \excitedcascadesixteen, \excitedcascadetwenty, and their \cc states.
	\subsubsection*{\textbf{Final States Particles}}
	The reconstruction of the charged final states particles is similar to the reconstruction presented in in the previous analysis.
	In addition, the neutral candidate list is filled whenever hits in the EMC cannot be associated with any charged track.
	Not using PID information leads to large combinatorics in the reconstruction process. Therefore, various selection criteria are used as a pre-filter for the candidates to reduce this combinatorics. 
	The track filtering is already described in \cref{ssec:trackfilter}.
	In addition to the track filter, the PID information is used as veto. 
	The PID value is calculated by using information about the energy loss $dE/dx$ in the detector material, the Cherenkov angle and the EMC cluster energy.\\
	Proton and antiproton candidates, which have a PID probability of more than $90\,\%$ to be a pion, are excluded.
	The same is applied for pions with a PID probability of more than $90\,\%$ to be a proton.
	The achieved reconstruction efficiency of the charged final state particles is summarized in \cref{tab:RecoEff_FS_Albrecht}.
	\begin{table}[htbp]
		\centering
		\caption{Reconstruction efficiency of the charged final state particles. The statistical error is on the order of $0.05\,\%$}
		\label{tab:RecoEff_FS_Albrecht}
		\begin{tabular}{lr}
			\hline
			Particle & Efficiency [\%] \\
			\hline
			\piminus(\lam) & 78.63 \\
			\piminus(\cascade) & 83.89 \\
			\piplus(\alam) & 78.67 \\		
			\piplus(\anticascade) & 84.07 \\
			p & 96.52 \\
			\aprot & 93.21 \\
			\hline
		\end{tabular}
	\end{table}
	For a further reduction of combinatorics, the candidates are subject to kinematical constraints on the transversal versus longitudinal momentum ($P_{t}$ vs. $P_{z}$) distribution. The elliptic boundary of the kinematically allowed region is given by
	\begin{equation}
		\frac{\left(x-x_0\right)^2}{a^2}+ \frac{y^2}{b^2} = 1
	\end{equation}
	with
	\begin{eqnarray*}
		x_0 &=& \left(p_{z,\max}+p_{z,\min}\right)/2 \\
		a &=& \left(p_{z,\max}-p_{z,\min}\right)/2, \mt{ and} \\
		b &=& p_{t,\max}.
	\end{eqnarray*}
	In this analysis, an event is marked as reconstructable, if the event contains the minimum number of entries according to the charged final states, \fsalbrecht as well as two neutral candidates.
	\subsubsection*{\textbf{Intermediate State Particles}}
	\label{ssec:intermediateStatesAlbrecht}
	The first step is to reconstruct the \lam and \alam particles.
	For \lam the list of protons and \piminus candidates are combined, for \alam those of \aprot and \piplus.
	Apart from this the procedure for \lam and \alam are identical.
	If not otherwise stated, the following description for \lam applies to the \alam reconstruction in the same way.\\
	The \lam candidates are first filtered by requiring that the $\pi^-$p mass ($M_{\mt{raw}}$) is within the following range: $\left|M_{\mt{raw}}-M_\Lambda\right|<0.15$\massunit.
	Here the lower bound of the mass window is given by the sum of the masses of the daughter particles.
	At this stage of reconstruction it is possible to reconstruct $30.5\,\%$ of the generated \lam and about $29.4\,\%$ of the generated \alam.\\
	\newline
	In order to reconstruct \cascade and \anticascade \mbox{(anti-)} hyperons, candidate pairs of \piminus and \lam or \piplus and \alam are built, respectively.
	The pion candidates from the respective candidate lists, which where used for the reconstruction of \lam and \alam, are excluded.
	Unless otherwise stated, the description of the \cascade reconstruction implicitly includes the reconstruction of \anticascade as well.
	In principle, the same procedure as for the \lam and \alam reconstruction is used.
	In a first step, the \cascade candidates are  filtered by a coarse mass window $\left|M_{\mt{raw}}-M_{\Xi^-}\right|< 0.15$\massunit, where the lower bound of the mass window is given by the sum mass of the daughter particles $M_\Lambda + M_{\pi^-}$.
	At this stage of the reconstruction, the reconstruction efficiency is $27.9\,\%$ for \cascade and $27.0\,\%$ for \anticascade.\\
	\newline
	The procedure to reconstruct the \pinull meson differs from the procedure for the hyperons.\\
	In the first step of the reconstruction, all members in the neutral candidates list are required to have at least $15\unit{MeV}$.
	To improve the \pinull selection, a photon time cut is introduced to reject neutrons. 
	For each neutral candidate a flight time difference of $T -T_{\mt{v=c}} < 3\unit{ns}$ is required, where $T$ is the recorded time of the first hit in the EMC.\\
	All pairwise combinations from the neutral candidate list are entered into the \pinull candidate list if the invariant mass of the pair ($M_{\mt{cand}}$) is within the following coarse mass window: $\left|M_{\mt{cand}}-m_{\pi^0}\right| < 0.05$\massunit with $m_{\pi^0} = 134.9768\unit{MeV}/c^2$ \cite{PDG2018} is then applied to these candidates.
	All candidates are subject to a mass constraint fit.
	A minimum fit probability threshold of $10^{-3}$ is required.
	If more than one candidate passes the fit, the candidates with the highest and second highest fit probability are selected.
	We separately counted MC truth \pinull decays into two photons whereby one or both of the photons have converted into a $e^+e^-$  pair in the material in front of the EMC.
	Therefore, the sum of true and \enquote{conversion} \pinull candidates is counted as good candidates leading to a fraction for \pinull signal events of $40.2\,\%$.
	The remaining candidates can be interpreted as combinatorial background.
	\subsubsection*{\textbf{Reconstruction of the $\mathbf{\bar{\Xi}^+\Xi^-\pi^0}$ System}}
\begin{figure}[h]
		\begin{subfigure}{0.5\textwidth}
			\centering
			\includegraphics[width=0.8\textwidth]{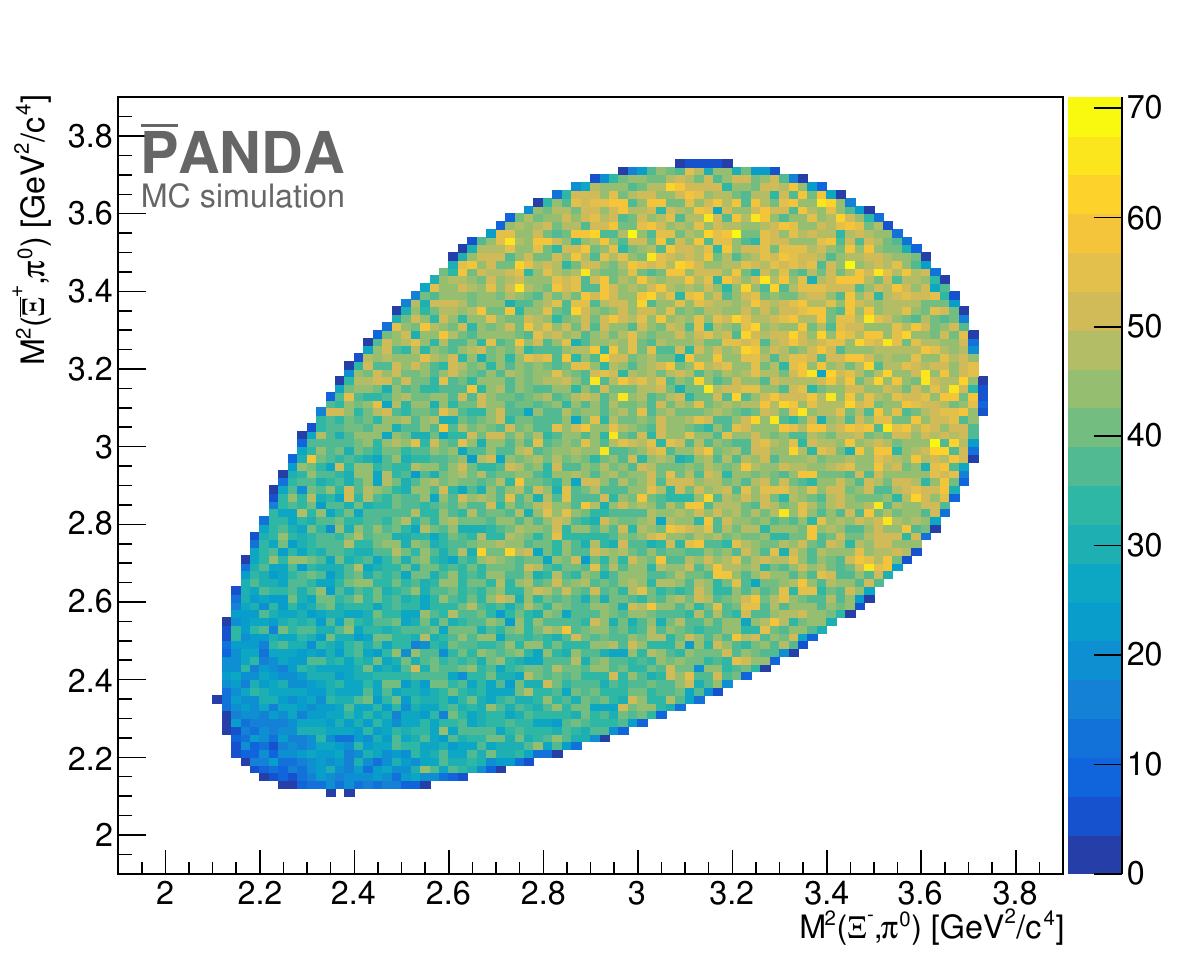}
			\caption{}
			\label{fig:DalitzPlot_reco_cont}
		\end{subfigure}
		\begin{subfigure}{0.5\textwidth}
			\centering
			\includegraphics[width=0.8\textwidth]{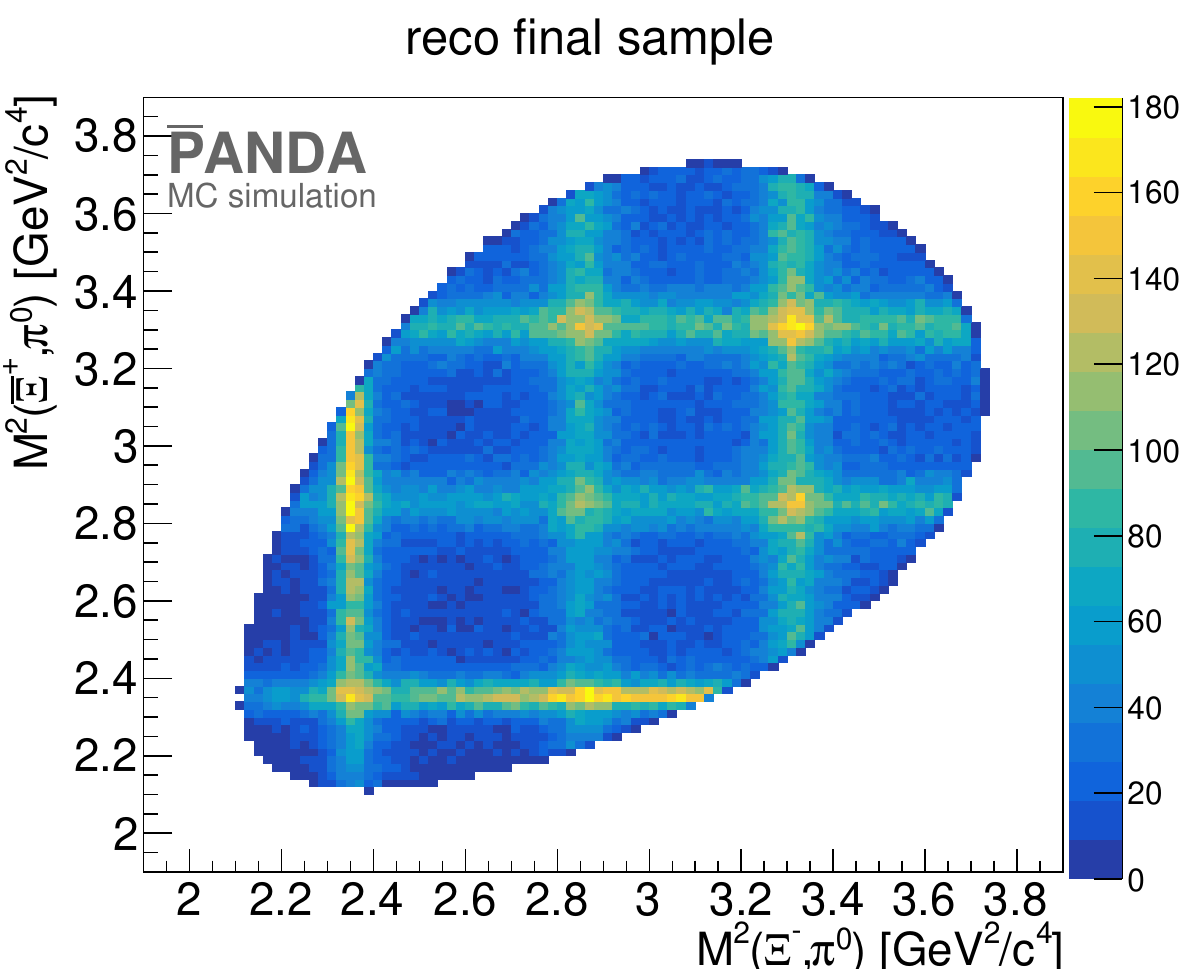}
			\caption{}
			\label{fig:DalitzPlot_reco_res}
		\end{subfigure}
		\caption{Dalitz plot for the final selection \cascasbarpinull candidates from the continuum contribution only (a) and for the resonance contribution only (b).}	
		\label{fig:DalitzPlots_Albrecht}
	\end{figure}

\begin{figure}[h]
		\centering
		\includegraphics[width=0.49\textwidth]{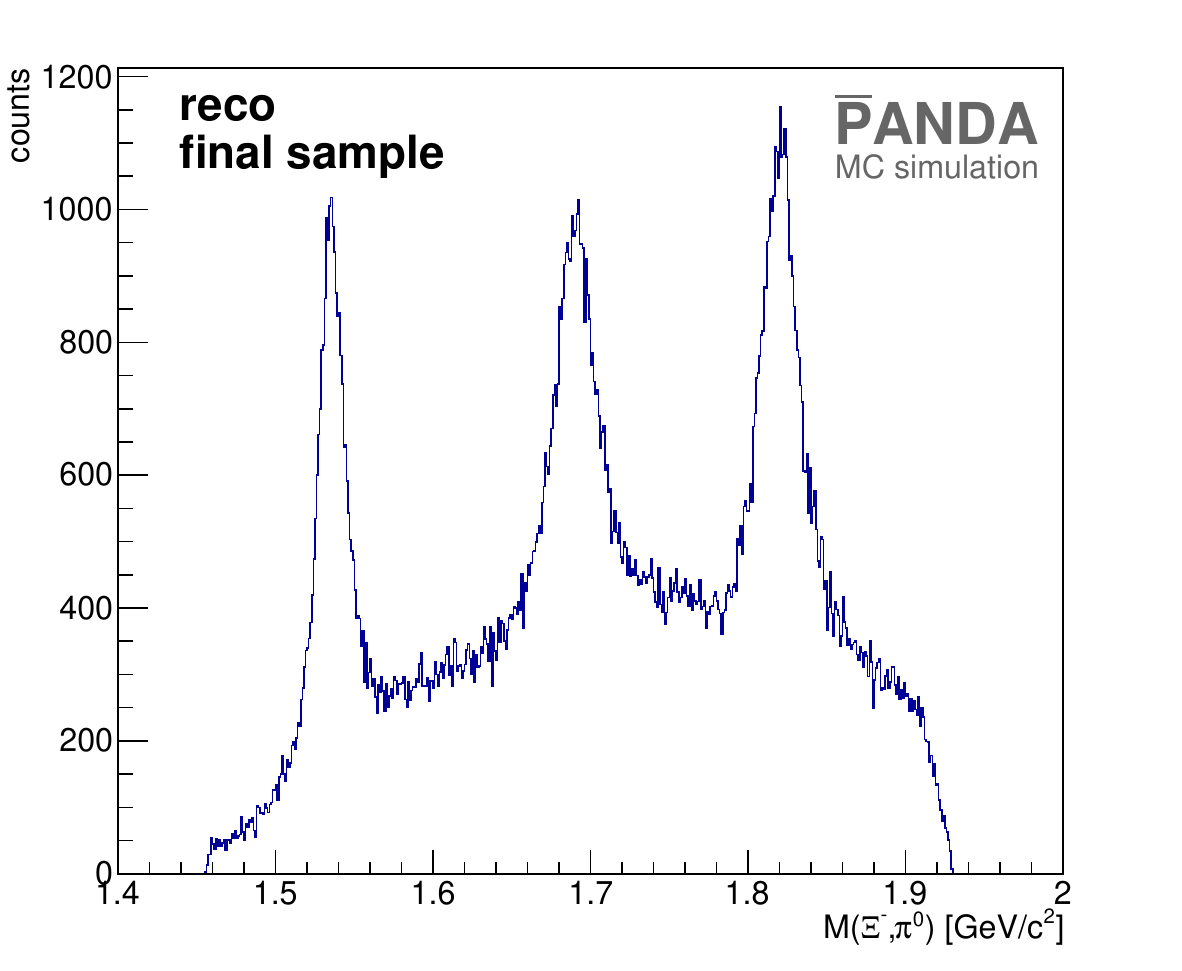}
		\caption{Mass distribution of the final selected \cascade\pinull sub-system.}
		\label{fig:MassDist_Reco_final_XiP0}
	\end{figure}

	In the last step of the analysis, the complete \cascasbarpinull system is combined.
	The combination of the three particles leads to a high amount of combinatorics.
	To reduce the number of \enquote{accidental} combined candidates a selection on the momentum in each component is performed corresponding to a selective cut on the four-momentum of the initial \pbarp system:
	\begin{eqnarray*}
		-0.14\unit{GeV/c}  < & P_{x,y} & <  0.14\unit{GeV/c}\\
		4.2\unit{GeV/c}  < & P_{z\,}  &<  5.0\unit{GeV/c} \\
		5.3\unit{GeV} < & E & < 5.9\unit{GeV}\\
		3.155\unit{GeV/c}^2 < & M & < 3.35\unit{GeV/c}^2.
	\end{eqnarray*}
	All remaining candidates are then subject to a full decay tree fit.
	In addition to the standard fits (vertices, four-momentum, masses), the constraint of the hyperon masses and \pinull mass are required.\\
	The fit results showed that the mass constraint of the \pinull is not perfectly fulfilled.
	To reduce the number of the candidates with a mass different from $M_{\pi^0}=0.135$\massunit \cite{PDG2018} the decay tree fit is redone with a corrected energy component for the \pinull candidates with too low masses.
	Finally, a minimum fit probability threshold of more than $10^{-4}$ is required to select the candidate.
	The probability threshold was chosen according to reach the best figure of merit in terms of reconstruction efficiency and pure signal fraction of the final selected sample.
	The described selection scheme leads to a reconstruction efficiency of $3.6\,\%$. 
	The most significant losses occur in the reconstruction of \pinull mesons.
	The signal purity of the final selected \cascasbarpinull candidates is $93.5\,\%$.
	In order to estimate the reconstructed signal event rate, the number of remaining signal events are multiplied by the product of all branching fractions of $0.4026$ within the decay tree, the luminosity and the cross section.\\
	The Dalitz plots for the final selected \cascasbarpinull are shown in \cref{fig:DalitzPlots_Albrecht}.
	In case of the continuum contribution, shown in \cref{fig:DalitzPlot_reco_cont}, the distribution differs from an expected uniform distribution.
	A loss of efficiency towards low $\Xi\pi^0$ masses is observable.
	The reason for the efficiency loss has to be investigated in the future.
	Nevertheless, the loss of efficiency is smooth so that this Dalitz plot could be analyzed.
	The contributing resonances are clearly observable as bands in \cref{fig:DalitzPlot_reco_res}.
	As an example, the mass distribution of the final selected \cascade \pinull sub-system is shown in \cref{fig:MassDist_Reco_final_XiP0}.
	\cref{tab:MassWidth_Reco_final_XiP0} summarized the obtained masses and widths of the contributing resonances by fitting the single peaks.
	In this study, the chosen input value for the \excitedcascadetwenty mass as well as the width of \excitedcascadesixteen and \excitedcascadetwenty were slightly different compared to te former study.
	The determined resonance masses are in good agreement with the input values, while the width for all resonances deviate from the input.
	Nevertheless, the fit values for the $\Xi$ and $\bar{\Xi}$ resonances are consistent with each other.
	\begin{table}[h]
		\caption{Fit results of for the mass and width of the $\Xi$ resonances determined with a fit to the peaks in the \cascade\pinull and \anticascade\pinull invariant mass distribution.}
		\label{tab:MassWidth_Reco_final_XiP0}
		\begin{tabular}{lcc}
			\hline
			& M [$\mt{MeV}/\mt{c}^2$] & $\Gamma$ [$\mt{MeV}/\mt{c}^2$]\\\hline
			\excitedcascadefifteen & $1535.9 \pm 0.3$ & $10.4 \pm 0.4$ \\
			\excitedanticascadefifteen & $1536.0 \pm 0.3$ & $10.4 \pm 0.4$ \\
			\excitedcascadesixteen & $1690.4 \pm 0.2$ & $21.7 \pm 0.5$ \\
			\excitedanticascadesixteen & $1690.7 \pm 0.2$ & $21.1 \pm 0.5$ \\
			\excitedcascadetwenty & $1819.8 \pm 0.3$ & $20.1 \pm 0.7 $\\
			\excitedanticascadetwenty & $1820.3 \pm 0.3$ & $20.5 \pm 0.7$\\
			\hline
		\end{tabular}
	\end{table} 
	\section{Background Studies}
	\label{sec:BackgroundStudies}
	In addition to the study of the signal channel, a study of hadronic background events is performed.
	The most critical contribution to background are processes ending in similar final states, e.g. \pbarp$\rightarrow \mt{p}\bar{\mt{p}}\pi^+\pi^+\pi^-\pi^-\mt{K}^+\mt{K}^-$ for \mychannelfs and \pbarp$\rightarrow \mt{p}\bar{\mt{p}}\pi^+\pi^+\pi^-\pi^-\pi^0$ for \channelalbrecht.
	In the latter case, the cross section is estimated to be on the order of $100\,\mu\mt{b}$ by extrapolating the results from \cite{Flaminio1984}. 
	Here, data samples were generated with the Dual Parton Model \cite{Capella1994} based generator DPM \cite{Galoyan2005} including only inelastic processes.
	The DPM event generator simulates all possible hadronic reactions for a given beam momentum.
	The cross-section of the \pbarp process is parameterized based on experimental data.\\ 
	100 million background events were subject to the same analysis strategy used for the signal events.
	In case of \mychannelfs, no event out of these 100 million background events survived the analysis procedure.\\
	In the study of $\bar{\mt{p}}\mt{p}\rightarrow$\cascasbarpinull, 7 events remained in the event sample after applying the full analysis procedure.
	Further studies showed, that these events could be removed by restricting the distance between the \cascade and \anticascade decay vertices $d_{\Xi - \bar{\Xi}}$.
	By requiring $d_{\Xi - \bar{\Xi}} > 1\unit{cm}$, the signal reconstruction efficiency is reduced to $3.1\,\%$.\\
	The non-observation of background events corresponds to a $90\,\%$ confidence upper limit of 2.3 events, which is used to calculate a lower limit for the signal-to-background ratio as well as for the signal significance.
	The signal-to-background ratio is given by
	\begin{equation}
		\frac{S}{B} = \frac{\sigma_{\mt{sig}}\cdot \epsilon_{\mt{sig}}\cdot b_{\mt{sig}}}{\sigma_{\mt{bg}}\cdot \epsilon_{\mt{bg}}},
	\end{equation}
	where $\sigma_{\mt{sig}}$ and $\sigma_{\mt{bg}}$ are the signal and inelastic \pbarp cross sections, respectively, $b_{\mt{sig}}$ is the total branching ratio of signal events, and $\epsilon_{\mt{sig}}$ and $\epsilon_{\mt{bg}}$ are the respective reconstruction efficiencies for signal and background.
	Since the signal cross sections has not yet been measured, for the \fs signal final state including also the continuum contribution, it is assumed to be $\sigma_{\mt{sig}}=1\,\mu\mt{b}$ and for \cascasbarpinull to be $2\,\mu\mt{b}$, since in experimental studies the cross section for the \cascasbarpinull ground state was determined to be higher than the cross section for the \fs ground state \cite{Musgrave1965}.
	Furthermore, the inelastic \pbarp cross section at a beam momentum of $4.6\momentumunit$ is $\sigma_{\mt{bg}}=50\unit{mb}$ \cite{PDG2018}.
	During the generation of the signal events, the branching ratio of \lam and \alam was set to $100\,\%$ for the decay \mbox{\decay{\lam}{p}{\piminus}} and \mbox{\decay{\alam}{\aprot}{\piplus}}. 
	For the following calculations this ratio has been  corrected by the factor $b_{\mt{sig}}=b^2_\Lambda = 0.4083$ for both final states investigated here.\\
	For the signal events, the reconstruction efficiency is $\epsilon_{\mt{sig}}=5.4\,\%$ for both \fs and \fscc, and $\epsilon_{\mt{sig}}=3.37\,\%$ for \cascasbarpinull.
	The significance of the signal $S_{\mt{sig}}$ is given by
	\begin{equation}
		S_{\mt{sig}} = \frac{N_{\mt{sig}}}{\sqrt{N_{\mt{sig}}+N_{\mt{bg}}\cdot F_{\mt{bg}}}},
		\label{eq:Significance}
	\end{equation}
	where $F_{\mt{bg}}$ denotes a scaling factor which corrects the number of background events according to the number of signal events, since the generated ratio for signal and background does not reflect the cross sections.
	The scaling factor is given by
	\begin{equation}
		F_{\mt{bg}} = \frac{N_{\mt{sig}}^{\mt{gen}}\cdot \sigma_{\mt{bg}}}{N_{\mt{bg}}^{\mt{gen}}\cdot \sigma_{\mt{sig}}\cdot b_{\mt{sig}}},
		\label{eq:ScalingFactor}
	\end{equation}
	where $N_{\mt{sig}}^{\mt{gen}}$ and $N_{\mt{bg}}^{\mt{gen}}$ are the number of generated signal and background events, respectively.
	\cref{eq:Significance,eq:ScalingFactor} transform to 
	\begin{equation}
		S_{\mt{sig}} = \frac{\sqrt{N^{\mt{gen}}_{\mt{sig}}}\cdot \epsilon_{\mt{sig}}}{\sqrt{\epsilon_{\mt{sig}}+\frac{\epsilon_{\mt{bg}}\cdot \sigma_{\mt{bg}}}{\sigma_{\mt{sig}}\cdot b_{\mt{sig}}}}}.
	\end{equation}
	In to following, the signal significance is calculated with the expected number of events within 3 days of data taking.
	This is motivated by the beam time which is need to collect the statistics necessary for a future partial wave analysis.
	Assuming a luminosity of $L=10^{31}\unit{cm}^{-2}\mt{s}^{-1}$, $\sigma_{\mt{sig}}=1\,\mu\mt{b}$ for \fs and $\sigma_{\mt{sig}}=2\,\mu\mt{b}$ for \cascasbarpinull, the expected number of events is $N^{\mt{gen}}_{\mt{sig}}\approx 12\cdot 10^6$ for \fs as well as for the \cc channel, and $N^{\mt{gen}}_{\mt{sig}}\approx 24\cdot 10^6$ for \cascasbarpinull.
	The calculated signal-to-background ratio and signal significance for each investigated channel are summarized in \cref{tab:SBandSignificance}.
	We also included the results based on a factor 10 smaller cross section to give an indication of the lower limit case.
	\begin{table}[htb]
		\centering 
		\caption{Signal-to-background ratio and signal significance. In addition to the assumed cross sections, calculations for a cross section of a factor 10 less are done.}
		\label{tab:SBandSignificance}
		\begin{tabular}{llcc}
			\hline
			& $\sigma_{\mt{sig}}$ & \fs ($\&$c.c.) & \cascasbarpinull\\
			\hline
			$S/B$ & $\sim 1\,\mu\mt{b}$  & $>19.1$ & $>22.0$\\
			$S_{\mt{sig}}$ & $\sim 1\,\mu\mt{b}$ & $>361$ & $>392$ \\
			$S/B$ & $\sim 0.1\,\mu\mt{b}$  & $>1.91$ & $>2.2$\\
			$S_{\mt{sig}}$ & $\sim 0.1\,\mu\mt{b}$ & $>95$ & $>105$ \\
			\hline
		\end{tabular}
	\end{table}
	%
	\section{Results and Discussion}
	In the previous section the feasibility study of the reactions  \mychannelfs, \mychannelfscc, and \channelalbrecht was described.\\ 
	In absence of experimental data and theoretical predictions for the angular distribution of the signal events, a uniform phase space distribution was assumed.
	This assumption is reasonable, since the amount of energy above the threshold is low for both channels and both strange valence quarks have been pair produced from the sea. 
	Here, this simplification assures, that the produced \cascade and \anticascade hyperons are underlying the same detector acceptance.
	An ideal pattern recognition was used for the track reconstruction in both analyses, since a realistic tracking algorithm for secondary tracks is currently not available.
	Therefore, a track filter was introduced to make the charged final state particle selection more realistic.\\
	The single particle reconstruction efficiency for the charged final state particles is between $68\,\%$ and $96\,\%$. 
	The intermediate state particles are reconstructed by applying a coarse mass window symmetrically around the nominal hyperon mass.
	With the resulting candidates, the three-body systems \fs, \fscc, and \cascasbarpinull are reconstructed and fitted with the DecayTreeFitter..	
	In the analysis of \fs and \fscc, a reconstruction efficiency of $\sim 5\,\%$ is achieved for each channel while for \cascasbarpinull a reconstruction efficiency of $3.6\,\%$ is achieved.	
	The obtained sample purity is $97.7\,\%$ for both \fs, and \fscc and $93.5\,\%$ for \cascasbarpinull, implying that the genealogy of the signal is suppressing the combinatorial background efficiently.\\
	The decay tree includes six final state particles in case of \mychannelfs (+ c.c.) and eight for \channelalbrecht.
	Here, the combined acceptance of the final state particles is limiting the reconstruction efficiency.
	In the study of \channelalbrecht the most limiting factor is the reconstruction of \pinull$\rightarrow \gamma\gamma$, since the reconstruction efficiency for \pinull is only about $40\,\%$.
	An improvement of the neutral particle reconstruction will also improve the reconstruction of the the \pinull candidates.\\
	With the assumed cross section of $1\,\mu\mt{b}$ for each considered final state, \fs and \fscc, the determined reconstruction efficiencies and the initial luminosity of $L=10^{31}\unit{cm}^{-2}\mt{s}^{-1}$, the expected reconstructed number of events is 38,500 per day.
	For the \cascasbarpinull final state a cross section of $2\,\mu\mt{b}$ is assumed.
	With the corresponding reconstruction efficiency and the initial luminosity, 22,800 reconstructed events are expected per day.
	These rates correspond to about 15 days of data taking to collect data samples with the same size of the reconstructed samples shown in this report.\\
	As already indicated, the reconstruction efficiencies as well as the expected number of reconstructed events depends on the assumed track efficiency of the tracking algorithm.
	Based on the results of the standard tracking algorithm, a track efficiency of about $90\,\%$ can be assumed reducing the signal reconstruction efficiency for \fs from $\sim5\,\%$ to $3\,\%$. 
	The selected events will be used as input for a partial wave analysis (PWA) of the \fs final state. 
	Here, one of the most important results is that there are no acceptance holes for the reconstructed sample.
	The statistics used to perform a partial wave analysis of a three-body final state is in the order of magnitude between 1,000 and 100,000 events \cite{ablikim2013,ablikim2013b,munzer2018}.
	From the results presented in \cite{Puetz2020} as well as from the follow up study for which a paper is currently in preparation, it is expected that for the \fs final state 30,000 reconstructed events are needed.
	The signal reconstruction efficiency of $3\,\%$ corresponds to about 10,500 reconstructed events per day at the initial luminosity and will give the possibility to collect the needed statistics for a PWA within a few days of data taking.\\
	For the study of the hadronic background the same analysis strategies were used as for the signal, leading to no surviving event out of 100 million generated background events for the \fs and \fscc final states.
	For the \cascasbarpinull final state seven events survived the applied cuts.
	Additional selection based on the distance between the \anticascade and \cascade vertices removed all background, but also reduced the overall signal efficiency to $3.1\,\%$.
	The background studies showed that at a $90\,\% $ confidence level a signal-to-background ratio of $S/B>19.1$ for \fs, $S/B>19.5$ for \fscc and $S/B>22$ for \cascasbarpinull could be achieved.
	The lower limit for the signal significance is $S_{\mt{sig}}>364$ for \fs, $S_{\mt{sig}}>361$ for \fscc and $S_{\mt{sig}}>392$ for \cascasbarpinull.
	To further quantify the signal-to-background ratio and the signal significance for \cascasbarpinull, future studies
	have to be performed with at least a factor 10 larger background sample.
	From the limits that we obtained, we can already conclude that it is feasible to produce a clean data sample necessary to perform a partial wave analysis.
	\newline
	Both analyses demonstrate that the experimental study of the process \mychannelfs, its \cc channel and \mbox{\channelalbrecht}, including also resonant baryon states, is feasible with the \panda detector.
	\section{Summary and Outlook}
	A first step has been done in investigating the feasibility of studying the $\Lambda$K and the $\Xi\pi$ decay of $\Xi$ resonances with the \panda detector in the reaction \mychannel and its \cc channel at an antiproton beam momentum of $4.6\momentumunit$.\\
	In the \fs study, a reconstruction efficiency of about $5\,\%$ has been achieved with a sample purity of $98\,\%$.
	The total reconstruction efficiency corresponds to 277,133 \fs events and 283,617 \fscc events.
	Assuming an initial luminosity $L=10^{31}\unit{cm}^{-2}\mt{s}^{-1}$, that number of final selected signal events can be collected within 15 days of data taking.
	100 million generated DPM background events were subject to the same selection strategy. 
	No background event survived, so that on a $90\,\%$ confidence level a lower limit for the signal significance of 361 for \fs and 392 \fscc has been determined.\\
	In the analysis of the \cascasbarpinull signal events the obtained total reconstruction efficiency is $3.6\,\%$, before selecting the (anti-)hyperon decay vertex position with respect to the interaction point.
	The sample purity of the final selected sample is $\sim93\,\%$. 
	The fake combinations in the sample are dominated by accidental combinations of neutral candidates in the reconstruction of the \pinull mesons.
	The total reconstruction efficiency of the signal events corresponds to about $3.2\cdot10^5$ events which can be collected in 15 days of data taking at the luminosity of $10^{31}\unit{cm}^{-2}\mt{s}^{-1}$.
	The identical analysis of 100 million DPM background events results in seven events surviving the applied cuts. 
	These events can be removed by requiring a separation of more than $1\unit{cm}$ between the \cascade and \anticascade decay vertex.
	The additional restrictions reduce the signal reconstruction efficiency to $3.1\,\%$.
	A lower limit for the signal-to-background ratio is deduced to be larger than 22, and the signal significance to be larger than 392.\\
	\newline
	The discussion in the previous chapter shows various steps that should be included in the analyses presented in the future.
	One point refers to the usage of the ideal tracking algorithm.
	As soon as a realistic tracking algorithm for secondary particles is available, the results of both studies need to be confirmed.
	The second point is the selection of the final state particles.
	The impact of the various PID selection criteria on the total reconstruction efficiency, the sample purity as well as on the signal-to-background ratio and the signal significance should be investigated.
	Furthermore, the model dependency of the background events should be reduced by comparing the results to the output of the background generators.\\
	A major goal of the $\Xi$ spectroscopy program at \panda is the determination of the spin and parity quantum numbers of the $\Xi$ states. 
	Therefore, a partial wave analysis (PWA) of the reconstructed three-body has to be performed.
	First investigations on a PWA tool which can be combined with a PandaRoot simulation and analysis are ongoing \cite{Puetz2020}.
	\section*{Acknowledgements}
	\label{cha:acknowledgements}
	\addcontentsline{toc}{chapter}{Acknowledgements}
	
	We acknowledge financial support from
	the Bhabha Atomic Research Centre (BARC) and the Indian Institute of Technology Bombay, India;
	the Bundesministerium f\"ur Bildung und Forschung (BMBF), Germany;
	the Carl-Zeiss-Stiftung 21-0563-2.8/122/1 and 21-0563-2.8/131/1, Mainz, Germany;
	the Center for Advanced Radiation Technology (KVI-CART), Groningen, Netherlands;
	the CNRS/IN2P3 and the Universit\'{e} Paris-Sud, France;
	the Czech Ministry (MEYS) grants LM2015049, CZ.02.1.01/0.0/0.0/16 and 013/0001677, Czech Republic;
	the Deutsche Forschungsgemeinschaft (DFG), Germany;
	the Deutscher Akademischer Austauschdienst (DAAD), Germany;
	the European Union's Horizon 2020 research and innovation programme under grant agreement No 824093;
	the Forschungszentrum J\"ulich, Germany;
	the Gesellschaft f\"ur Schwerionenforschung GmbH (GSI), Darmstadt, Germany;
	the Helmholtz-Gemeinschaft Deutscher Forschungszentren (HGF), Germany;
	the INTAS, European Commission funding;
	the Institute of High Energy Physics (IHEP) and the Chinese Academy of Sciences, Beijing, China;
	the Istituto Nazionale di Fisica Nucleare (INFN), Italy;
	the Ministerio de Educación y Ciencia (MEC) under grant FPA2006-12120-C03-02, Spain;
	the Polish Ministry of Science and Higher Education (MNiSW) grant No. 2593/7, PR UE/2012/2, and the National Science Centre (NCN) DEC-2013/09/N/ST2/02180, Poland;
	the State Atomic Energy Corporation Rosatom, National Research Center Kurchatov Institute, Russia;
	the Schweizerischer Nationalfonds zur F\"orderung der Wissenschaftlichen Forschung (SNF), Switzerland;
	the Science and Technology Facilities Council (STFC), British funding agency, Great Britain;
	the Scientific and Technological Research Council of Turkey (TUBITAK) under the Grant No. 119F094, Turkey;
	the Stefan Meyer Institut f\"ur Subatomare Physik and the \"Osterreichische Akademie der Wissenschaften, Wien, Austria;
	the Swedish Research Council and the Knut and Alice Wallenberg Foundation, Sweden.
%

\bibliographystyle{ieeetr}
\bibliography{literature}

\end{document}